\documentclass[12pt,preprint]{aastex}

\usepackage{xcolor}

\shorttitle{Horizontal shear instabilities at low Prandtl number}
\shortauthors{Garaud et al.}

\begin{document}


\title{Horizontal shear instabilities at low Prandtl number}



\author{P. Garaud$^1$}
\affil{
$^{1}$Department of Applied Mathematics, Baskin School of Engineering, \\
University of California, Santa Cruz, CA 95064\\
}



\begin{abstract}
Turbulent mixing in the radiative regions of stars is usually either ignored or crudely accounted for in most stellar evolution models. However, there
is growing evidence that such mixing is present and can affect various aspects of a star's life. Here, we present a first attempt at quantifying mixing by horizontal shear instabilities in stars using Direct Numerical Simulations. The shear is driven by a body force, 
and rapidly becomes unstable. At saturation, we find that several distinct dynamical regimes exist, depending on the relative importance of 
stratification and thermal diffusion. In each of the regimes identified, 
we propose a certain number of theoretically motivated scaling laws for the turbulent vertical eddy scale, the turbulent diffusion coefficient,
and the amplitude of temperature fluctuations (among other quantities). Based on our findings, we 
predict that the majority of stars should fall into one of two categories: high P\'eclet number stratified turbulence, and low P\'eclet number stratified turbulence. 
The latter is presented in a related paper by Cope et al. (2020), while the former is discussed here. Applying our results to the solar tachocline, we find that it should lie in the high P\'eclet number stratified turbulence regime, and predict a substantial amount of
vertical mixing for temperature, momentum and composition. Taken as is, the new turbulence model predictions are incompatible with the Spiegel \& Zahn (1992) 
model of the solar tachocline. However, rotation and magnetic fields are likely to affect the turbulence, and need to be taken into account in future 
studies.
\end{abstract}



\section{Introduction}
\label{sec:intro}

Inspired by the seminal work of Jean-Paul Zahn on shear instabilities in stars \citep{Zahn1974,Zahn92}, we have begun a systematic exploration of this process and of its impact on mixing in stellar radiative zones \citep[see][]{Garaudal15,GaraudKulen16,Garaudal17,GagnierGaraud2018,KulenGaraud2018,Copeal20}.
Shear is almost omnipresent in stellar interiors. It can be directly measured in the Sun and in Red Giant Branch (RGB) stars thanks to helio- and astero-seismology \citep[see][and many others.]{JCDSchou88,Brownal89,Thompson-etal96,Schou-etal98,Charbonneaual99,Beck-etal11,Deheuvels12,Beck12,Mosser12a,Mosser12b,Deheuvelsetal14,Benomar2018,Bazot2019}. It can also be inferred from observations of the surface differential rotation of intermediate-mass stars \citep[][]{Barnes2005,Reiners2006,Reinhold2013,Balona2016}. Shear instabilities have long been invoked as a source of turbulent mixing in stars, participating in the transport of both angular momentum and chemical elements. Since the source of the shear is usually the star's differential rotation, shear-induced mixing is one of the many processes involved in what stellar astrophysicists usually refer to as rotational mixing \citep[see, e.g.][]{Zahn1974,Pinsonneault97,MaederMeynet2000}.

Shear can have components in both the vertical direction (radial shear) and in the horizontal direction (latitudinal shear), as exemplified by observations of the solar tachocline \citep[e.g.][]{Schou-etal98,Charbonneaual99}. Nevertheless, the vast majority of theoretical studies of shear-induced mixing to date have focussed on the effect of vertical shear only. Vertical shear instabilities are perhaps the most intuitive source of vertical mixing in stars, since they directly generate vertical fluid motion. But they are also directly affected by stratification, which tends to suppress vertical flows. Indeed, a parcel of fluid, displaced adiabatically by a distance $\Delta r$ from its original position $r_m$ would experience a buoyancy force
\begin{equation}
{\bf F}_b = \Delta r \left( \left. \frac{\partial \rho_0}{\partial r}\right|_{r=r_m} - \left. \frac{\partial \rho}{\partial r} \right|_{ad} \right) {\bf g}= - \rho_m \Delta r  N^2 {\bf e}_r, 
\end{equation}
where $\rho_0(r)$ is the background radial density profile of the star, $\rho_m = \rho_0(r_m)$, ${\bf g}$ is gravity, $\left. \partial \rho/\partial r \right|_{ad}$ is the rate of change of density a parcel would undergo while traveling adiabatically, and $N$ is the Brunt-V\"ais\"al\"a frequency. As such, the larger the stratification (as quantified by $N$), the larger the restoring force experienced by adiabatic motions. Stratification can suppress adiabatic vertical shear instabilities entirely unless the shear $S$ exceeds a certain threshold, such that 
\begin{equation}
J = \frac{N^2}{S^2} \le J_c, 
\end{equation}
where $J_c$ is a constant of order unity. This criterion is known as the Richardson criterion \citep{Richardson1920,Howard61}. The quantity $J$ is fundamental to the study of stratified vertical shear flows, and is the so-called gradient Richardson number. Typical values of $J$ in stellar interiors are usually in excess of $10^3$ even in very strong shear layers, suggesting that shear instabilities are not possible. However, \citet{Zahn1974} noted that thermal diffusion can be very large in stars \citep[see also][]{SpiegelZahn70}, so the displacement of fluid parcels is not necessarily adiabatic, especially if the latter are small. He then argued that the correct criterion to apply should instead be 
\begin{equation}
J Pr  \le (JPr)_c,
\end{equation}
where $Pr = \nu / \kappa_T$ is the Prandtl number (which is the ratio of the viscosity $\nu$ to the thermal diffusivity $\kappa_T$) and $(JPr)_c$ is also a constant, which is now $O(10^{-3})$. The validity of this criterion was independently verified by various groups \citep{PratLignieres13,PratLignieres14,GaraudKulen16,Pratal2016,Garaudal17} who further established that $(JPr)_c \simeq 0.007$.  Since the Prandtl number is usually exceedingly small in stars (being typically $\sim 10^{-6}$ or less), this implies that vertical shear instabilities can be excited even when the gradient Richardson number $J$ is very large (i.e up to $O(10^4)$ or so, depending on the local value of $Pr$). Thermal diffusion being important for these instabilities to develop, they are now commonly referred to as {\it diffusive} shear instabilities, or sometimes {\it secular} shear instabilities. 

Despite this, there are several reasons why these so-called diffusive vertical shear instabilities may not be a particularly important source of mixing in stars. First, even with the modified stability criterion proposed by \citet{Zahn1974}, vertical shear in most stars remains stable because the stratification is so strong (i.e. $N$ is very large); typical Richardson numbers in RGB stars for instance are $~O(10^5-10^6)$ assuming that the angular velocity profile is smoothly varying between the rapidly rotating core and slowly rotating envelope \citep{Beck-etal11,Deheuvels12,Deheuvelsetal14}. Second, the typical vertical eddy scale associated with diffusive shear instabilities is small, because it has to allow for rapid thermal diffusion; as proposed by \citet{Zahn92} and confirmed by the Direct Numerical Simulations (DNSs) of \citet{Garaudal17}, this scale is given by 
\begin{equation}
l_{\rm Z} =  \sqrt{\frac{\kappa_T S}{N^2} }  ,
\end{equation}
which we call the Zahn scale hereafter. Consequently, the corresponding turbulent diffusivity is also relatively small. \citet{Zahn92} suggested that it can be modeled as 
\begin{equation}
D_{\rm turb} \propto S l_{\rm Z}^2 \propto \frac{\kappa_T S^2}{N^2} = C \frac{ \kappa_T}{J},
\label{eq:dturb1}
\end{equation}
which was recently confirmed by \citet{PratLignieres14}, \citet{Pratal2016} and \citet{Garaudal17}, as long as it is applied in the correct parameter regime intended by \citet{Zahn92} \citep[see][for more detail]{Garaudal17}.   \citet{Garaudal17} estimated the constant $C$ to be around $0.08$, which would imply
\begin{equation}
D_{\rm turb} \simeq 8 \left(\frac{J}{10^5}\right)^{-1} \left( \frac{\kappa_T}{10^7 {\rm cm^2 / s}} \right) {\rm cm}^2 / {\rm s}.
\label{eq:dturb2}
\end{equation}
From (\ref{eq:dturb1}) we see that whenever $J \gg 1$, $D_{\rm turb} \ll \kappa_T$; this is expected since the instability only occurs because of strong thermal diffusion, so one would not expect it to transport heat faster than diffusely. However, we also see from (\ref{eq:dturb2}) that for the typical parameter values adopted here, $D_{\rm turb}$ is not much larger than the typical microscopic viscosity or compositional diffusivity in the star either (which are of the order of unity in cgs units). This implies, as stated above, that diffusive vertical shear instabilities may not a particularly relevant source of mixing for stellar evolution. 

An alternative source of shear-induced mixing, also discussed by \citet{Zahn92}, are horizontal shear instabilities. By contrast with vertical shear instabilities, which must necessarily involve vertical fluid motions, horizontal shear instabilities can develop with purely horizontal flows and are therefore unaffected by stratification. As such, they are always present except when stabilized by rotation \citep[see][]{Watson1980,Garaud01}. Note that without any vertical flow, purely horizontal shear instabilities cannot induce any advective vertical transport. However, \citet{Zahn92} further argued that the horizontal fluid motions in each radial shell could become decoupled, therefore leading to the generation of substantial vertical shear on short lengthscales. This would drive secondary diffusive vertical shear instabilities, and associated turbulent mixing. \citet{Zahn92} \citep[see also][for an alternative argument leading to the same scaling]{Lignieres2018} argued that the turbulent mixing coefficient associated with these horizontal flows would be
\begin{equation}
D_{\rm turb} \propto \sqrt{\frac{\kappa_T \varepsilon}{N^2}} ,
\end{equation}
(see his equation 2.22) where $\varepsilon$ is the viscous dissipation rate, which he assumes would be of the order of the mechanical energy injection rate. \citet{Copeal20} performed the first study of horizontal shear instabilities with a stellar context in mind, and confirmed Zahn's prediction for the turbulent diffusion coefficient, albeit only in a specific region of parameter space (see more on this issue in Section \ref{sec:Cope} below). It is interesting to note that 
$D_{\rm turb}$ now scales as $(\kappa_T / N^2)^{1/2}$, by contrast with the turbulent mixing coefficient associated with vertical shear instabilities, which scales as $\kappa_T / N^2$. As such, the former is more likely to dominate in the strong stratification limit than the latter. For this reason, we now propose to perform a more comprehensive study of mixing by horizontal shear instabilities in stars, building on the work of \citet{Zahn92} and \citet{Copeal20}. Note that preliminary results on this work were presented in \citet{Garaud20}, but our theoretical interpretation of the data has since changed, so the conclusions presented in this paper should be preferred.

Section \ref{sec:model} presents the setup used for our numerical experiments on horizontal shear instabilities, which is identical to that of \citet{Copeal20}. Section \ref{sec:Cope} summarizes the results of \citet{Copeal20} and clarifies why a more comprehensive study is needed. Section \ref{sec:num} describes the numerical method used, and analyzes the results both qualitatively and quantitatively. In particular,  \ref{sec:scalings} tentatively proposes a new model for mixing by stratified horizontal shear instabilities in stars, that should be valid in a wide range of parameter space. Finally, Section \ref{sec:disc} summarizes the results, discusses implications for the solar tachocline, and raises a number of further questions that need to be addressed before the model can safely be used in stellar evolution codes.

\section{Model setup}
\label{sec:model}

Following \citet{Copeal20} we consider a small region of the radiative zone of a star, located around radius $r_m$. Since we are ignoring the effects of rotation, the latitude of that region is irrelevant. We 
use a local Cartesian domain with coordinates $(x,y,z)$, where gravity is aligned with the vertical axis, so ${\bf g} = -g{\bf e}_z$. Here $z = r - r_m$ where $r$ is the local radius, $x$ is in the azimuthal direction, and $y$ is in the latitudinal direction. We use the \citet{SpiegelVeronis1960} Boussinesq approximation for weakly compressible gases, which is valid as long as the height of the computational domain $L_z$ is smaller than any density or temperature scaleheight, an assumption that is fairly reasonable deep in the stellar interior. Consistent with this assumption, the gravity $g$, viscosity $\nu$ and thermal diffusivity $\kappa_T$ are assumed to be constant within the domain. The background temperature profile $T_0(z)$ is assumed to be in thermal equilibrium, which then implies that the background temperature gradient $T_{0z}$ must be constant as well, within the context of the model used. As such, we have 
\begin{equation}
T_0(z)=T_m+T_{0z}z, 
\end{equation}
where $T_m$ is the mean temperature of the star near $r = r_m$.
Consistent with the Spiegel-Veronis-Boussinesq approximation, we assume that the equation of state can be linearized around $T_m$, such that
\begin{equation}
\rho_0(z) = \rho_m +  \left. \frac{\partial \rho}{\partial T}\right|_{T_m} T_{0z} z  = \rho_m \left( 1 - \alpha  T_{0z} z \right), 
\end{equation}
where $\rho_m = \rho(p_m,T_m)$ is the mean density of the region, which defines the coefficient of thermal  expansion $\alpha$ as the thermodynamic derivative
\begin{equation}
\alpha = - \rho_m^{-1}  \left. \frac{\partial \rho}{\partial T}\right|_{T_m}.
\end{equation}

A body force ${\bf F}$ is assumed to drive a mean shear flow in this domain, which in turn drives the development of shear instabilities. Perturbations to the background temperature arising from these instabilities are assumed to be triply-periodic in the domain, and the total temperature profile is
\begin{equation}
T(x,y,z,t) =  T_0(z) + \tilde{T}(x,y,z,t). 
\end{equation}
The linearized equation of state them implies that corresponding density perturbations can be written as
\begin{equation}
\frac{\tilde{\rho}}{\rho_m} = -\alpha \tilde{T}. \label{eq:eos}
\end{equation}

With these definitions in mind, the Spiegel-Veronis-Boussinesq equations governing the fluid evolution under the effect of a body-force ${\bf F}$ are:
\begin{eqnarray}
\frac{\partial{\bf u}}{\partial t} + {\bf u}\cdot\nabla {\bf u} & = & - \frac{1}{\rho_m} \nabla \tilde{p} + \alpha \tilde{T} g {\bf e}_z + \nu \nabla^2 {\bf u} +  \frac{1}{\rho_m} {\bf F} \label{eq:momentum}, \\
\nabla \cdot {\bf {u}} &=& 0 \label{eq:continuity},  \\
\frac{\partial \tilde{T}}{\partial t} + {\bf u}\cdot\nabla \tilde{T} + w  (T_{0z} - T_{ad,z})  & = & \kappa_T \nabla^2 \tilde{T} \label{eq:heat}, 
\end{eqnarray}
where ${\bf u} = (u,v,w)$ is the velocity field, $\tilde p$ is the pressure perturbation away from hydrostatic equilibrium (both ${\bf u}$ and $\tilde p$ are also assumed to be triply-periodic), $T_{ad,z} = - g/c_p$ is the adiabatic temperature gradient, and $c_p$ is the specific heat at constant pressure. For simplicity, we assume that the shear is driven by a sinusoidal body force ${\bf F} = F_0 \sin(k_s y) {\bf e}_x$ where $k_s  = 2\pi/L_y$ is the wavenumber associated with the domain width $L_y$. The mean flow is therefore in the $x$ (azimuthal) direction, while the mean shear is in the $y$ (horizontal / latitudinal) direction. 

As in \citet{Copeal20}, we now non-dimensionalize the variables and equations using the anticipated amplitude of the flow $U$, obtained by requiring a balance between the inertial terms and the forcing in the $x$ direction: 
\begin{equation}
({\bf u} \cdot \nabla) u \sim \rho_m^{-1} F_0  \sin(k_s y) \Rightarrow k_s U^2 \sim \rho_m^{-1} F_0 ,  
\end{equation}
which defines
\begin{equation}
U = \left( \frac{F_0}{\rho_mk_s} \right)^{1/2} 
\label{eq:Udef}
\end{equation}
as the unit velocity. The unit length is taken to be $k_s^{-1}$, so the unit time is $(k_s U)^{-1}$. Finally, we choose $k_s^{-1} (T_{0z} -  T_{ad,z}) $ as the unit temperature so the non-dimensional equations are 
\begin{eqnarray}
\frac{\partial \hat{\bf u}}{\partial t} + \hat{\bf u}\cdot\nabla \hat{\bf u} & = & - \nabla \hat{p} + B \hat{T} {\bf e}_z + Re^{-1} \nabla^2 \hat {\bf u} +   \sin(y) {\bf e}_x \label{eq:nondimmomentum}, \\
\nabla \cdot \hat{\bf u} &=& 0 \label{eq:nondimcontinuity},  \\
\frac{\partial \hat{T}}{\partial t} + \hat {\bf u}\cdot\nabla \hat{T} + \hat{w}  & = & Pe^{-1} \nabla^2 \hat{T} \label{eq:nondimheat}, 
\end{eqnarray}
where all the hatted quantities are from here on non-dimensional\footnote{To simplify the notation, we have not added hats on the independent variables $x,y,z$ and $t$, or on the differential operators; their non-dimensionalization is implicit.} and where
\begin{equation}
Re = \frac{U}{k_s \nu}, \quad Pe = \frac{U}{k_s \kappa_T} , \quad \mbox{ and } B = \frac{N^2}{k_s^2 U^2},  
\end{equation}
are the Reynolds number, P\'eclet number, and stratification parameters, respectively. $N$ is the Brunt-V\"ais\"al\"a frequency discussed earlier, defined here in terms of the quantities introduced so far as $N^2 = \alpha g (T_{0z} -  T_{ad,z})$. The Reynolds number is the usual ratio of the viscous diffusion timescale to the turbulent advection timescale, the P\'eclet number is the corresponding ratio of the thermal diffusion timescale and the turbulent advection timescale, and finally, $B$ is the square of the ratio of the buoyancy frequency to the shearing rate, and is the equivalent of the Richardson number but for horizontal shear.

Typical values of $Re$, $Pe$ and $B$ in stars can be estimated as follows, noting that $k_s = 2\pi/ L_y$ where $L_y$ is the dimensional lengthscale of the horizontal shear (which is presumed to be of the order of the stellar radius).  
\begin{eqnarray}
Re = \frac{10^{14}}{2\pi} \left(\frac{U}{10^4 {\rm cm/s}} \right) \left(  \frac{L_y}{10^{11} {\rm cm}} \right) \left( \frac{\nu}{10{\rm cm^2/s}}  \right)^{-1},  \nonumber \\
Pe = \frac{10^{8}}{2\pi}  \left(\frac{U}{10^4 {\rm cm/s}} \right) \left(  \frac{L_y}{10^{11} {\rm cm}} \right) \left( \frac{\kappa_T}{10^7{\rm cm^2/s}}  \right)^{-1}, \nonumber \\
B = \frac{10^{8}}{4\pi^2}  \left(\frac{U}{10^4 {\rm cm/s}} \right)^{-2} \left(  \frac{L_y}{10^{11} {\rm cm}} \right)^2 \left( \frac{N}{10^{-3}{\rm s}^{-2} } \right)^2.
\label{eq:tachopars}
\end{eqnarray}
We therefore see that for the usual stellar parameters selected here, $Re,Pe, B\gg 1$, while $Pr = Pe / Re \ll 1$. It is worth noting, however, that in the envelopes of high mass stars $\kappa_T$ can exceed $10^{15}$cm$^2$/s \citep{Garaudal15b}, in which case $Pe < 1$, as already noted by \citet{GaraudKulen16} (although $Re$ and $B$ remain much greater than 1). 

 The linear stability properties of horizontal sinusoidal shear flows in a vertically-stratified medium have been studied extensively  \citep{Lucasal2017,Copeal20}. Similar studies for a hyperbolic tangent shear profile were presented by \citet{AroboneSarkar2012} and \citet{Parkal2020}. The main findings of these studies are two fold. First, assuming that the domain is longer than wide (i.e. $L_x > L_y$), then the two-dimensional (2D), vertically-invariant mode of instability is always the most rapidly growing mode (provided $Re$ is larger than a factor of order unity). The properties of this 2D mode are independent of stratification ($B$) or thermal diffusion ($Pe$). Second, three-dimensional perturbations (i.e perturbations that vary with $z$) are also almost always excited, but their growth rates are often much smaller than that of the 2D mode. As demonstrated by \citet{Copeal20}, however, these 3D perturbations play a crucial role in the saturation of the instability in low P\'eclet number flows, and are responsible for the layerwise decoupling of the 2D modes central to the \citet{Zahn92} model for mixing by horizontal shear instabilities.

\section{Horizontal shear instabilities at low P\'eclet number}
\label{sec:Cope}

The simulations of \citet{Copeal20} focussed on a distinguished limit of these equations, namely the low P\'eclet number limit ($Pe < 1$). This section summarizes their results for completeness; the reader is referred to the original paper for more detail. 

The low P\'eclet number limit is interesting for two reasons. First, as noted above, this limit is indeed
achieved in the outer layers of high mass stars. Second, it lends itself to an asymptotic simplification of the governing equations, that greatly facilitates their analysis. Indeed, as
shown by \citet{Lignieres1999} \citep[see also][]{Spiegel1962,Thual}, in the limit where the P\'eclet number based on the actual eddy scale and the rms velocity of the flow is small, the advection terms in the heat equation 
are negligible in comparison with the advection of the background temperature profile. As a result the dominant balance in the temperature equation is
 \begin{equation}
\hat{w}  \simeq Pe^{-1} \nabla^2 \hat{T} \label{eq:LPN}, 
\end{equation}
which can then be applied in the momentum equation to yield 
\begin{equation}
\frac{\partial \hat{\bf u}}{\partial t} + \hat{\bf u}\cdot\nabla \hat{\bf u}  =  - \nabla \hat{p} + BPe \nabla^{-2} \hat w {\bf e}_z + Re^{-1} \nabla^2 \hat {\bf u} +   \sin(y) {\bf e}_x. \label{eq:LPNmomentum}
\end{equation}
 Together, equation (\ref{eq:LPNmomentum}) and continuity form the low P\'eclet Number (LPN) equations, and 
can be solved self-consistently instead of (\ref{eq:nondimmomentum})--(\ref{eq:nondimheat}). We see that the only relevant governing parameters are now $Re$ and $BPe$, which reduces by one the dimension of parameter space to be explored. More interestingly, (\ref{eq:LPN}) reveals that the temperature field is slaved to the velocity field, which strongly constrains the allowable dynamics, and can also be used to help interpret the results. 

\citet{Copeal20} ran a number of simulations using both the normal equations at $Pe \le 1$, and the low P\'eclet number equations. The results of the latter were consistent with those of the former when run at the same values of $Re$ and $BPe$. Using these simulations, \citet{Copeal20} were able to identify four distinct dynamical regimes (two of which are only present for sufficiently high Reynolds numbers). In all cases, the initial development of the instability was consistent with predictions from linear theory: the vertically-invariant 2D mode is always the first to grow, followed by 3D perturbations that cause a vertical modulation of the 2D perturbations. This results in a series of meandering horizontal jets that are only weakly coupled in the vertical, and drive substantial vertical shear, as proposed by \citet{Zahn92}. In the limit of small stratification (low $BPe$), the buoyancy force is essentially negligible. The vertical and horizontal shear rapidly become fully turbulent, and the turbulence supports a continuous range of eddy scales from the injection scale (which here is $\hat L_y$) down to the viscous scale. This is the unstratified regime, where heat is merely a passive tracer. As $BPe$ increases above unity, the vertical shear instability between the meanders continues to exist, but the increasing stratification gradually reduces the vertical size of turbulent eddies and the vertical velocity. In the horizontal direction, the flow contains both large scales (associated with the forcing) and small scales (associated with the vertical eddy scale through continuity). Turbulence is present throughout the domain, which is an important characteristic of this turbulent stratified regime. As $BPe$ continues to increase, however, the vertical eddy scale becomes sufficiently small for viscosity to be important. This begins to affect (but does not entirely suppress) the vertical shear instability, and the turbulence becomes intermittent. Finally, at the largest values of $BPe$, the turbulence is entirely suppressed by viscosity and the flow dynamics become layerwise two-dimensional, with each thin layer viscously connected to its neighbors. 

 A quantity of particular interest for mixing in stratified fluids is the so-called mixing efficiency $\eta$ \citep[see, e.g.][]{Maffioli2016} which measures  how much of the energy injected into the system is dissipated thermally vs. viscously. Indeed, dotting the momentum equation with $\hat {\bf u}$ and integrating the result over the (periodic) domain yields the kinetic energy conservation equation
\begin{equation}
\frac{\partial}{\partial t} \langle \frac{|\hat {\bf u}|^2}{2} \rangle  = B\langle \hat w \hat T \rangle - Re^{-1} \langle | \nabla \hat {\bf u} |^2 \rangle + \langle \sin(y) \hat u \rangle ,  
\end{equation}
where $\langle \cdot \rangle$ denotes a volume average hereafter. Terms on the right-hand side are the rate of conversion of kinetic energy into potential energy, the viscous dissipation rate, and the mechanical energy input rate, respectively. 
Multiplying the temperature equation with $\hat T$ and integrating the result over the domain yields the potential energy conservation equation
\begin{equation}
\frac{\partial}{\partial t} \langle \frac{\hat T^2}{2} \rangle  = - \langle \hat w \hat T \rangle - Pe^{-1} \langle | \nabla \hat T |^2 \rangle , 
\end{equation}
where the second term on the right-hand side is the thermal dissipation rate. Assuming a statistically stationary state, and combining these equations, we get
\begin{equation}
\langle \sin(y)\hat u \rangle  = - B\langle \hat w \hat T \rangle + Re^{-1} \langle | \nabla \hat {\bf u} |^2 \rangle = BPe^{-1} \langle | \nabla \hat T |^2 \rangle + Re^{-1} \langle | \nabla \hat {\bf u} |^2 \rangle, 
\end{equation}
which indeed shows that the energy injected into the flow can be dissipated in two ways: viscously, or thermally. The quantity $\eta$ is then defined as
\begin{equation}
\eta = \frac{- B\langle \hat w \hat T \rangle}{ \langle \sin(y) \hat u \rangle  } ,
\label{eq:etadef}
\end{equation} 
and can be interpreted as the ratio of the amount of kinetic energy transferred to potential energy (and later dissipated thermally), to that injected into the flow mechanically by the force ${\bf F}$. 

Together with the qualitative observations summarized earlier, \citet{Copeal20} were able to model the dynamics of each of these regimes (other than the intermittent regime) using arguments of dominant balance, and proposed various scaling laws for the vertical eddy scale $\hat l_z$, the rms vertical velocity $\hat w_{rms}$, the rms temperature fluctuations $\hat T_{rms}$, and $\eta$. These are summarized in Table \ref{tab:scalings}. Combining the expression for $\hat w_{rms}$ and $\hat l_z$ yields a prediction for the vertical turbulent mixing coefficient $\hat D_{turb}$, also shown in Table \ref{tab:scalings}. We see that in the stratified turbulent regime, $\hat D_{turb}$ scales as $(BPe)^{-1/2}$, which can easily be shown to recover Zahn's model for mixing by horizontal shear flows. 

 \begin{table}[h]
        \caption{\small Scaling laws in the low P\'eclet number regime, as determined by \citet{Copeal20}. The prefactors are specific to the sinusoidal horizontal shear flow adopted here, but the scalings should be universally valid. The value of $\eta$ in the turbulent stratified regime should also be universally valid. The scalings in the intermittent regime are empirical only. }
        \centering{
        \vspace{0.3cm}
{\small
\begin{tabular}{ccccc}
                \tableline
                 Regime & Unstratified Turb. & Stratified Turb. & Intermittent & Viscous             \\
                        \tableline
Validity   & $BPe \ll 1$  &  $1 \ll BPe \ll 0.0016 Re^2$  & $0.0016 Re^2 \ll BPe \ll 5 Re^2$ & $BPe \gg 5 Re^2$  \\ 
                        \tableline
$\hat l_z$ & $2$  & $2(BPe)^{-1/3} $ & $2(BPe)^{-1/3} $ & $2Re^{-1/2} $  \\ 
$\hat w_{rms}$ &  $1$  & $ (BPe)^{-1/6} $ & $ 0.05Re^{3/4}(BPe)^{-1/2} $  & $ 0.25Re^{3/2}(BPe)^{-1} $ \\
$\hat T_{rms}$ & $Pe$ & $ Pe(BPe)^{-5/6}$ & $ Pe(BPe)^{-5/6}$  & $ PeRe^{1/2}(BPe)^{-1}$  \\
$\eta$ &  $ 0.4BPe$ & $0.4 $   & $0.08Re^{1/2}(BPe)^{-1/4} $  & $0.25Re^{2}(BPe)^{-1} $   \\
$\hat D_{turb}$ & $2$ & $2(BPe)^{-1/2}$ & $  0.1Re^{3/4}(BPe)^{-5/6}  $ & $  0.5Re (BPe)^{-1}  $ \\
 \tableline
\end{tabular}
}
}
\label{tab:scalings}
\end{table}

The scaling laws for the stratified turbulent regime are derived as follows \citep{Copeal20}. First, note that the horizontal component of the flow velocity $\hat u$ must be $O(1)$ by the non-dimensionalization selected. Next, since $Pe$ is low, we have from (\ref{eq:LPN}) and on dimensional grounds that 
\begin{equation}
\hat w_{rms} \sim Pe^{-1} \hat l_z^{-2} \hat T_{rms}.
\end{equation}
Assuming a balance in the vertical component of the momentum equation between the nonlinear term $\hat {\bf u} \cdot \nabla \hat w$ and the buoyancy term $B\hat T$, then we also have 
\begin{equation}
\hat u_{rms} \hat w_{rms} \hat l_z^{-1} \sim B\hat T_{rms}.
\end{equation}
Combining the two implies 
\begin{equation}
\hat l_z  \sim (BPe)^{-1/3}.
\end{equation}
Next, if one {\it assumes} that $\eta$ is roughly constant and of order unity (and take this as a defining property of this regime), then $- B\langle \hat w \hat T \rangle \sim \langle \hat u \sin(y) \rangle \sim O(1)$, so 
\begin{equation} 
B \hat w_{rms} \hat T_{rms} \sim O(1). 
\end{equation}
Combining this with the above, we then obtain
\begin{eqnarray}
\hat w_{rms} \sim (BPe)^{-1/6}, \hat T_{rms} \sim (BPe)^{-5/6} Pe, \\
\hat D_{turb} \sim \hat w_{rms} \hat l_z \sim (BPe)^{-1/2}.
\end{eqnarray}
It is worth noting that the theoretical derivation of this scaling law differs somewhat from the derivations of \citet{Zahn92} or \citet{Lignieres2018}, despite arriving at the same conclusion for $\hat D_{turb}$. This is because  \citet{Zahn92} and \citet{Lignieres2018} assume that the viscous dissipation is known and fixed, while we assume that the mechanical forcing (and therefore the typical horizontal flow velocity $U$) is known and fixed. However, the conclusions are consistent otherwise.

The applicability of the results of \citet{Copeal20} is limited to low P\'eclet number flows (using the P\'eclet number $Pe$ that is based on the large-scale properties of the shear), and since these conditions are only realized in the envelopes of the most massive stars, they should not a priori be used to model mixing in intermediate mass main sequence stars. In this paper we therefore extend their analysis to flows for which $Pe \gg 1$, but $Pr \ll 1$ (as is the case in the majority of stellar interiors). 

\section{Numerical simulations} 
\label{sec:num}

\subsection{Methodology}

As in \citet{Copeal20}, we use DNSs to investigate the nonlinear evolution of stably stratified horizontal shear flows. We use the pseudo-spectral PADDI code \citep{Traxleretal2011b,Stellmachetal2011}, modified to account for the presence of a body force \citep[e.g.][]{Garaudal15,GaraudKulen16,GagnierGaraud2018}, to solve equations (\ref{eq:nondimmomentum})-(\ref{eq:nondimheat}). The computational domain is triply-periodic, with size ($\hat L_x, \hat L_y, \hat L_z$). The dimensions $\hat L_x$, $\hat L_y$ and $\hat L_z$ are $4\pi$, $2\pi$ and $2\pi$, respectively, after \citet{Copeal20}. This selection was found to be a good tradeoff between computational feasibility and dynamical reliability, i.e. the ability to capture the correct dynamics without being overly affected by the boundary conditions \citep{CopeGFD}. The computational costs of these simulations is indeed high: since we focus in this paper on the high P\'eclet number and low Prandtl number regime, and since $Re = Pe / Pr$, the Reynolds number has to be very high, and the resolution of the simulations has to be correspondingly high as well. Furthermore, multiple simulations at high $Re$ are required to capture the parametric dependence of the solution on $Re$, $Pe$ and $B$. As such, we choose in what follows to focus on 2 series of simulations only:
\begin{itemize}
\item Simulations at $Pr = 0.1$, with $Re = 100$ ($Pe = 10$), $Re = 300$ ($Pr = 30$) and $Re = 600$ ($Pe = 60$). 
\item Simulations at $Pr = 0.05$, with $Re = 600$ ($Pe = 30$). 
\end{itemize}
Table \ref{tab:runs} presents all the available runs, together with selected salient properties. All simulations at $Re = 100$ have a resolution of $384 \times 192 \times 192$ equivalent grid points; those at $Re = 300$ have $576 \times 288 \times 288$ equivalent grid points, and finally those at $Re = 600$ have $768 \times 384 \times 384$ equivalent grid points. The adequacy of the resolution was checked for each simulation by visual inspection of the energy spectrum, of the physical space vorticity field, and by computing the product of the Kolmogorov scale and of the largest wavenumber (which needs to be greater than one). 

Simulations were either started from initial conditions with $\hat u(x,y,z,0) = \sin(y)$ and all other fields seeded with random small amplitude perturbations, or, from another simulation at nearby parameters (e.g. gradually increasing or decreasing $B$). We have found that the initial conditions used have no influence on the nature of the statistically stationary state reached by the simulation, whenever such a state is achieved. However, it is not always easy to be certain that such a state has been reached, especially for simulations at large $B$ and $Re$ (which are computationally expensive). Details of the issues arising are presented in Appendix A. Generally speaking, we find that quantities associated with vertical transport (such as the rms vertical velocity and the rms temperature perturbations) very rapidly reach a stationary state, and in all the cases presented in Table \ref{tab:runs} such a state has indeed been achieved. However, quantities associated with horizontal transport (such as the rms horizontal velocities) sometimes exhibit variability on very long timescales in the limit of large stratification. Table \ref{tab:runs} lists which simulations have not reached a statistically stationary state in terms of horizontal transport, and which have. 
 
 \begin{table}[]
        \caption{\small Parameters and main results for the high P\'eclet number DNSs. The fourth column shows $\hat U_{rms}$ (see equation \ref{eq:Urms}), the fifth, sixth and seventh show $\hat w_{rms}$, $\hat T_{rms}$, and $\eta$, respectively (see Section \ref{sec:dataextract}) and the last column shows the vertical lengthscale $\hat l_{z2}$ (see Appendix B). All measurements are taken as time averages once the system has reached a statistically stationary state, and the $\pm$ represents the rms variability around the mean. In simulations for which $\hat U_{rms}$ has not reached a steady state but all other quantities have (see Appendix A), $\hat U_{rms}$ is written in brackets.}
        \centering{
        \vspace{0.3cm}
        {\small 
\begin{tabular}{cccccccc}
                \tableline
                
$Re$ & $Pe$ & $B$ & $\hat U_{rms}$ & $\hat w_{rms}$ & $\hat T_{rms}$  &$\eta$ & $\hat l_{z2}$  \\
                        \tableline
 100 & 10 & 10 & 2.73  $\pm$ 0.22 & 0.52  $\pm$   0.10 & 0.18 $\pm$  0.03 & 0.38  $\pm$ 0.03 & 1.26 $\pm$ 0.10 \\ 
 100 & 10 & 30 & 2.56  $\pm$ 0.22 & 0.19  $\pm$   0.04 & 0.06 $\pm$  0.008 & 0.22  $\pm$ 0.04 & 0.87 $\pm$ 0.08 \\ 
 100 & 10 & 100 & 2.10  $\pm$ 0.08 & 0.05  $\pm$   0.004 & 0.018 $\pm$  0.001 & 0.16 $\pm$ 0.02 & 0.49 $\pm$ 0.02 \\ 
 100 & 10 & 1000 & 2.21  $\pm$ 0.11 & 0.016  $\pm$   0.001 & 0.004 $\pm$  0.0005 & 0.10  $\pm$ 0.01 & 0.38 $\pm$ 0.03 \\ 
 100 & 10 & 10000 & 3.44  $\pm$ 0.46 & 0.004  $\pm$   0.002 & 0.0008 $\pm$  0.0003 & 0.04  $\pm$ 0.02 & 0.47 $\pm$ 0.12 \\ 
\tableline 
300	& 30	& 0.01 & 2.43 $\pm$ 0.12 & 0.96 $\pm$ 0.08 & 1.16 $\pm$ 0.24 & 0.01 $\pm$ 0.002 & 1.91 $\pm$ 0.16 \\
300	& 30 & 0.1 & 2.39 $\pm$ 0.11 &  0.94 $\pm$ 0.08 & 0.83$\pm$ 0.11 & 0.07 $\pm$ 0.01 & 1.77 $\pm$ 0.20 \\
300 & 30 & 1 &  2.29 $\pm$ 0.17 & 0.82 $\pm$ 0.10 & 0.55 $\pm$ 0.05 & 0.30 $ \pm$ 0.04 & 1.62 $\pm$ 0.20 \\
300 & 30 & 10  &2.50 $\pm$ 0.18 & 0.61 $\pm$ 0.08 & 0.21 $\pm$ 0.02 & 0.44 $\pm$ 0.04 & 1.03 $\pm$ 0.10 \\
300 & 30 & 30  &   3.13 $\pm$ 0.19 & 0.46 $\pm$ 0.09 & 0.10 $\pm$ 0.02 & 0.40 $\pm$ 0.03 & 0.75 $\pm$ 0.06 \\
300 & 30 & 100 & (3.19 $\pm$ 0.09) & 0.19 $\pm$ 0.05 & 0.038 $\pm$ 0.004 &  0.27 $\pm$ 0.04 & 0.60 $\pm$ 0.06 \\ 
300 & 30 & 300 & (3.22 $\pm$ 0.07) & 0.03 $\pm$ 0.009 & 0.011 $\pm$ 0.001& 0.17 $\pm$ 0.02 & 0.32 $\pm$ 0.03 \\ 
300 & 30 & 1000 & (4.08 $\pm$ 0.07) & 0.02 $\pm$ 0.003 & 0.0054 $\pm$ 0.0006 & 0.13 $\pm$ 0.02 & 0.22 $\pm$ 0.01 \\
300 & 30 & 10000 & (2.42 $\pm$ 0.07) & 0.004 $\pm$ 0.0003 & 0.0008 $\pm$ 5$\cdot 10^{-5}$ & 0.05 $\pm$ 0.004  &  0.18 $\pm$  0.008 \\
 \tableline 
600 & 30 & 10 & 2.15 $\pm$ 0.11 & 0.57 $\pm$ 0.05 & 0.19 $\pm$ 0.02 & 0.47 $\pm$ 0.02 & 1.03 $\pm$ 0.09 \\ 
600 & 30 & 30 & 2.40 $\pm$ 0.13 & 0.42 $\pm$ 0.06 & 0.09 $\pm$ 0.01 & 0.43 $\pm$ 0.02 & 0.63 $\pm$ 0.05 \\
600 & 30 & 100 & (2.56 $\pm$ 0.11) &  0.25 $\pm$ 0.04 & 0.036 $\pm$ 0.003 &   0.35 $\pm$ 0.03 &  0.43 $\pm$ 0.02 \\ 
\tableline
600 & 60 & 0.1 & 2.36 $\pm$ 0.16 & 0.94 $\pm$ 0.12   & 0.87 $\pm$  0.10 & 0.07 $\pm$ 0.01  &1.93 $\pm$ 0.29 \\
600 & 60 & 1 & 2.33 $\pm$ 0.19  & 0.86 $\pm$ 0.07 & 0.61 $\pm$ 0.05  &  0.28 $\pm$  0.04 & 1.69 $\pm$ 0.20 \\
600 & 60 & 10 & 2.13 $\pm$ 0.13 & 0.59  $\pm$ 0.06 & 0.21 $\pm$ 0.02 & 0.47 $\pm$  0.03 &  1.04 $\pm$ 0.08 \\
600 & 60 & 100 & 2.19 $\pm$ 0.10 & 0.26 $\pm$ 0.06  & 0.04 $\pm$  0.005 & 0.33 $\pm$  0.03 & 0.46 $\pm$ 0.04\\
600 & 60 & 400 & (2.99 $\pm$ 0.13) & 0.13 $\pm$ 0.06 & 0.015 $\pm$ 0.002 & 0.19 $\pm$ 0.05 & 0.36 $\pm$ 0.03 \\
600 & 60 & 1000 & (2.80 $\pm$  0.1) & 0.02 $\pm$ 0.01 & 0.006 $\pm$  0.0007 &  0.14 $\pm$ 0.02 & 0.20 $\pm$ 0.01 \\
600 & 60 & 6000 & (2.28 $\pm$ 0.11) & 0.007 $\pm$ 0.0006 & 0.002 $\pm$ 0.0002 & 0.12 $\pm$  0.02 & 0.17 $\pm$ 0.01 \\
\tableline
\end{tabular}
}
}
\label{tab:runs}
\end{table}

\subsection{Qualitative behavior of the flow}

\begin{figure}[h]
\epsscale{0.5}
\plotone{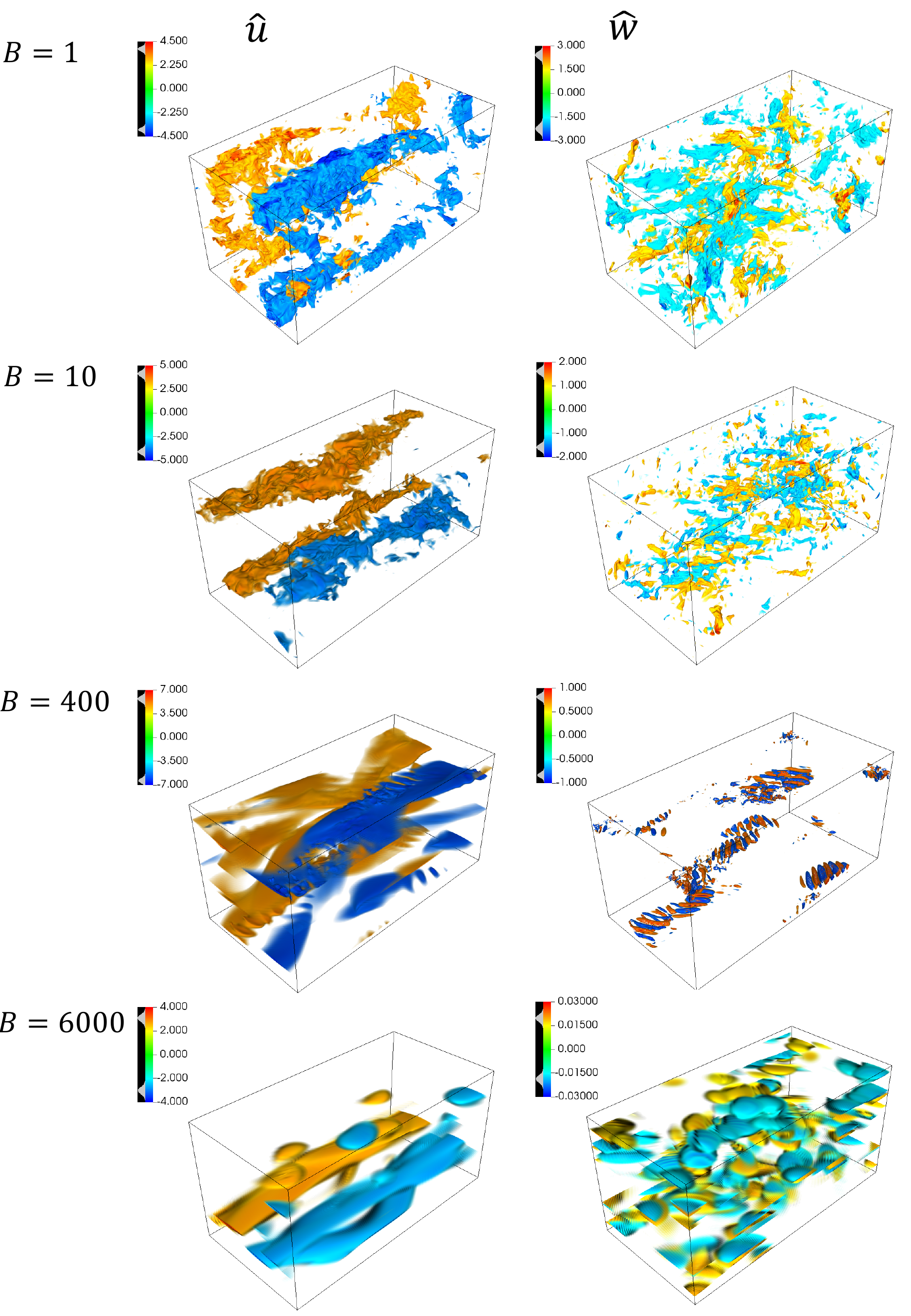}
\caption{Volume-rendered \citep{Visit} snapshots of $\hat u$ and $\hat w$ for simulations with $Re = 600$, $Pe = 60$, and varying $B$. From top to bottom, we see the unstratified regime, the stratified turbulent regime, the intermittent regime and the viscous regime.}
\label{fig:SnapshotsRe600}
\end{figure}

From a purely qualitative point of view, we find that properties of our simulations at high Reynolds number, high P\'eclet number and low Prandtl number are similar to those of high Reynolds number / low P\'eclet number flows. In particular, we find that they appear to be divided into the same four regimes identified by \citet{Copeal20}: an unstratified regime, a turbulent stratified regime, an intermittent regime, and a viscous regime. Volume-rendered snapshots of $\hat u$ and $\hat w$ in each regime, for simulations with $Re = 600$ and $Pe = 60$ (so $Pr = 0.1$) are shown in Figure \ref{fig:SnapshotsRe600}.  The unstratified regime (here, for $B = 1$) is qualitatively identical to that described by \citet{Copeal20}; this is not surprising, since the temperature field (not shown) behaves like a passive scalar in that limit. The turbulence exhibits a wide range of scales, from the domain scale to the viscous scale.  In the stratified turbulence regime (here for $B = 10$), turbulence is present everywhere in the domain as well, but the vertical eddy scale is smaller; the meanders of the horizontal flow are more clearly visible. In the intermittent regime (here for $B = 400$), as the name suggests, the turbulence is intermittent in both time and space. The eddy scale is even smaller, and is affected by viscosity; this can be seen by the fact that the vertical shear instability takes the form of much more organized and localized rolls. Finally, for very large values of $B$ (here, for $B = 6000$), the vertical shear instability is entirely viscously suppressed. The horizontal flow takes the form of thin meandering jets in each layer, and a very weak vertical flow is generated from the divergence of the horizontal flow. 

\subsection{Data extraction}
\label{sec:dataextract}

For all simulations presented in Table \ref{tab:runs}, we have measured the time-dependent quantity 
\begin{equation}
\hat q_{rms}(t) = \langle \hat q^2(x,y,z,t) \rangle^{1/2} ,
\end{equation}
where $\hat q$ could be $\hat u$, $\hat v$, $\hat w$ or $\hat T$. If that quantity has achieved a statistically stationary state, then we take the time average of $\hat q_{rms}(t)$ over the interval $\Delta t$ 
for which the system is statistically stationary, and report it in Table 2 as $\hat q_{rms}$, and the associated errorbar quantifies the rms time variability of $\hat q_{rms}(t)$ around $\hat q_{rms}$. Simulations for which a statistically stationary state has been reached for $\hat w_{rms}(t)$, $\hat T_{rms}(t)$ and $\hat \eta(t)$ but not for $\hat u_{rms}(t)$ and/or $\hat v_{rms}(t)$ (see discussion and example in Appendix A), are shown in brackets.

In all cases, we have also computed an estimate of the vertical eddy size, using the first zero of the vertical autocorrelation function of $\hat w$ \citep[see, e.g.][]{Garaudal17,Copeal20}. More specifically, we computed
\begin{equation}
A_w(\hat l,t) = \langle \hat w(x,y,z,t) \hat w(x,y,z+\hat l,t) \rangle ,
\label{eq:Aw}
\end{equation}
and let $\hat l_z(t)$ be the first zero of $A_w(\hat l,t)$. We then take the time average of $\hat l_z(t)$ over the duration of the statistically stationary state available, and the associated errorbar quantifies the rms time variability of $\hat l_z(t)$ around the mean $\hat l_z$. Note that this is done as a post-processing step for the simulations, and since the full fields are not stored very often, the computation of $\hat l_z$ does not always involve many instants in time.  

Finally, we compute the time-dependent mixing efficiency as
\begin{equation}
\eta(t) = \frac{ - B \langle \hat w \hat T \rangle }{- B \langle \hat w \hat T \rangle + Re^{-1} \langle |\nabla \hat {\bf u} |^2 \rangle } ,
\label{eq:etaoft}
\end{equation}
and report $\eta$ in Table \ref{tab:runs} as the time average of $\eta(t)$ during the statistically stationary phase, together with its rms variability.

\subsection{Quantitative results}

\begin{figure}
\epsscale{0.8}
\plotone{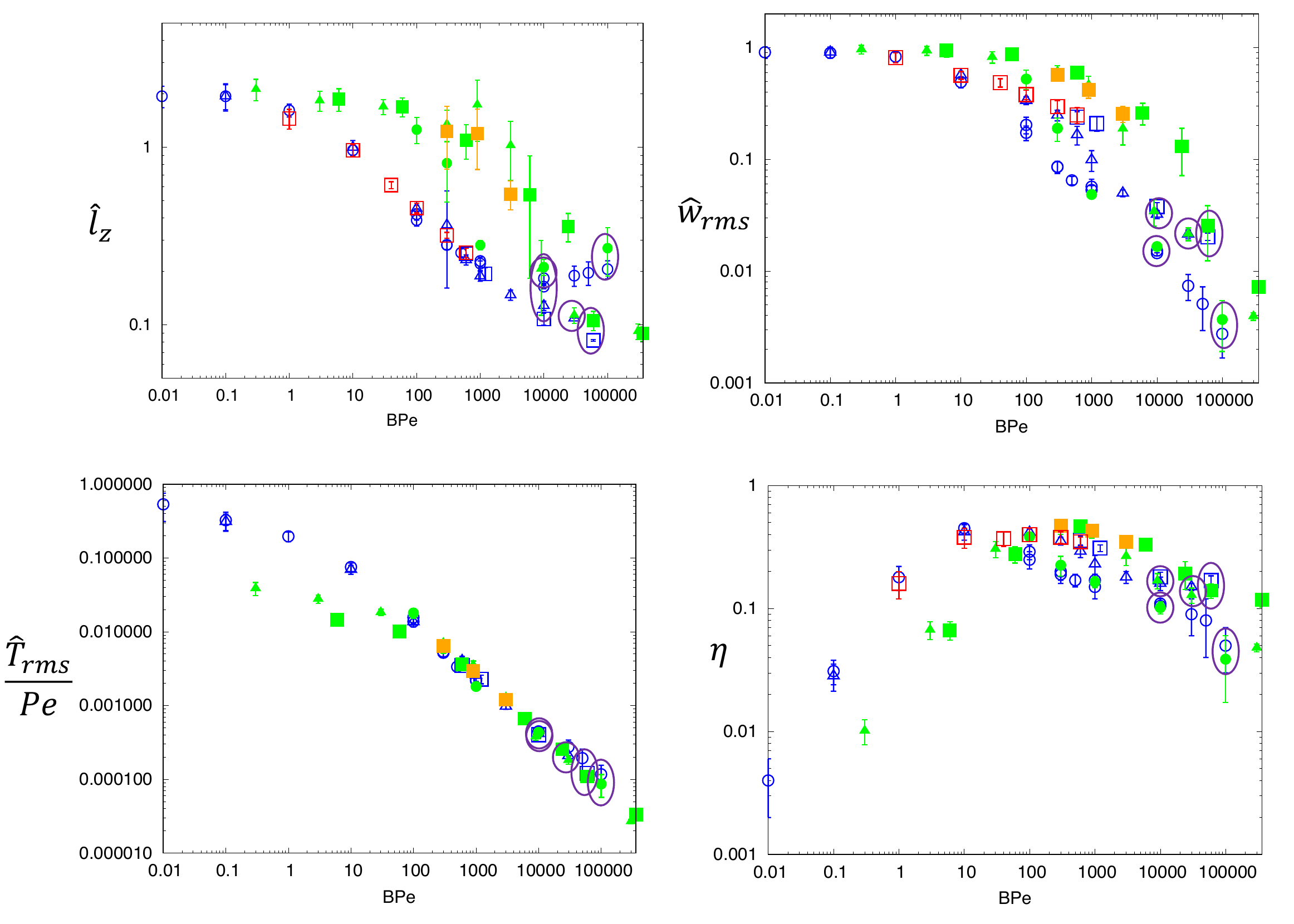}
\caption{From top left to bottom right: $\hat l_z$, $\hat w_{rms}$, $\hat T_{rms}/Pe$ and $\eta$ as functions of $BPe$. In each quadrant, blue symbols represent simulations at $Pe \le 1$; red symbols represent simulations performed using the LPN approximation; green symbols represent high $Pe$ simulations at $Pr = 0.1$ and orange symbols have $Pr = 0.05$. The shape/size of the symbol represents the Reynolds number: small circles for $Re = 100$, small triangles for $Re = 300$ and large squares for $Re = 600$. Simulations that have large $Pe$, but whose dynamics appear to satisfy LPN dynamics, and lie close to a low $Pe$ point at the same parameter values, are marked by a purple ellipse (see main text for detail).}
\label{fig:HighPe_results}
\end{figure}

\citet{Copeal20}, who focussed on the low P\'eclet number limit, presented all their quantitative results on the flow statistics as functions of $BPe$ and $Re$ (see their Figure 8). This is a natural choice for their data since these are the only two relevant parameters at low $Pe$ (see Section \ref{sec:Cope}). By contrast, there is no reason to expect that the flow statistics should only depend on $BPe$ and $Re$ in high P\'eclet number systems. Nevertheless, to ease the comparison of our results with those of \citet{Copeal20}, we first present them as functions of $BPe$ in Figure \ref{fig:HighPe_results}. In all cases, the shape / size of the symbol identify the Reynolds number (small circle for $Re = 100$, small triangle for $Re  =300$, and large square for $Re = 600$). Open symbols are used for the data presented by \citet{Copeal20}, with blue symbols corresponding to simulations using the normal equations (\ref{eq:nondimmomentum})-(\ref{eq:nondimheat}) with $Pe \le 1$, while red symbols correspond to simulations run using the asymptotic low P\'eclet number equation (\ref{eq:LPNmomentum}). Filled symbols are used to present the new data obtained for this paper; the green-colored symbols correspond to the suite of simulations with $Pr = 0.1$ and the orange-colored symbols correspond to $Pr =   0.05$. 
 
We clearly see from this comparison between the high P\'eclet and low P\'eclet number data that the qualitative similarity of the results discussed earlier does not translate into a quantitative similarity. On the whole, the high P\'eclet number data is quite distinct from the low P\'eclet number data. However, a closer inspection of Figure \ref{fig:HighPe_results} shows that a few points for $Pe \gg 1$ lie on top of (or very close to, and within the errorbars of) those at $Pe \ll 1$.  Crucially, these pairs of points have the same Reynolds number, and the same values of $BPe$, but have different individual values of $B$ and $Pe$. These points are marked with a purple ellipse and are generally located in the region of parameter space corresponding to the intermittent or viscous regimes. As we now demonstrate, this is not a coincidence.

\subsubsection{When does a flow exhibit low P\'eclet number dynamics?} 
\label{eq:LPNdynamics}

 As discussed by \citet{Lignieres1999} and summarized earlier, the condition that needs to be met to be in the asymptotically low P\'eclet number regime is not $Pe \ll 1$ (where we recall that $Pe$ is defined based on the outer scales of the system) but instead, $Pe_t \ll 1$, where $Pe_t$ is the {\it turbulent} P\'eclet number based on the actual flow velocities and actual eddy scale. Since the eddy scale decreases with increasing stratification, it is quite plausible that $Pe_t$ could drop below unity thus leading to low P\'eclet dynamics even when $Pe \gg 1$. This idea is in fact central to Zahn's model for horizontal shear instabilities \citep{Zahn92}, and was confirmed numerically by \citet{GaraudKulen16} for vertical shear instabilities.
 
 To test it here, we need a simple way to determine when a system is dominated by low P\'eclet dynamics (i.e. when $\hat w \simeq Pe^{-1} \nabla^2 \hat T$) and when it is not. One could compute at each point in the domain and each point in time the respective sizes of the terms $\hat w$, $\hat {\bf u} \cdot \nabla \hat T$ and  $Pe^{-1} \nabla^2 \hat T$, and compare them to one another; however, this is unnecessarily cumbersome. After analyzing various possibilities, we have determined that the ratio 
 \begin{equation}
r =  \frac{|\hat F_T| }{\hat w_{rms} \hat T_{rms}} , 
\label{eq:r}
 \end{equation}
 where $\hat F_T$ is the time average of $\langle \hat w \hat T \rangle$ during the statistically stationary state, 
 is an excellent diagnostic of the flow dynamics. Indeed, for truly low P\'eclet number flows, (\ref{eq:LPN}) holds so $\hat w$ and $\hat T$ are exactly in phase with one another. As a result, $r$ is very close to one. On the other hand, when (\ref{eq:LPN}) does not hold, $\hat w$ and $\hat T$ are generally not in phase, and $r$ drops below one. 
 
\begin{figure}[h]
\epsscale{0.9}
\plotone{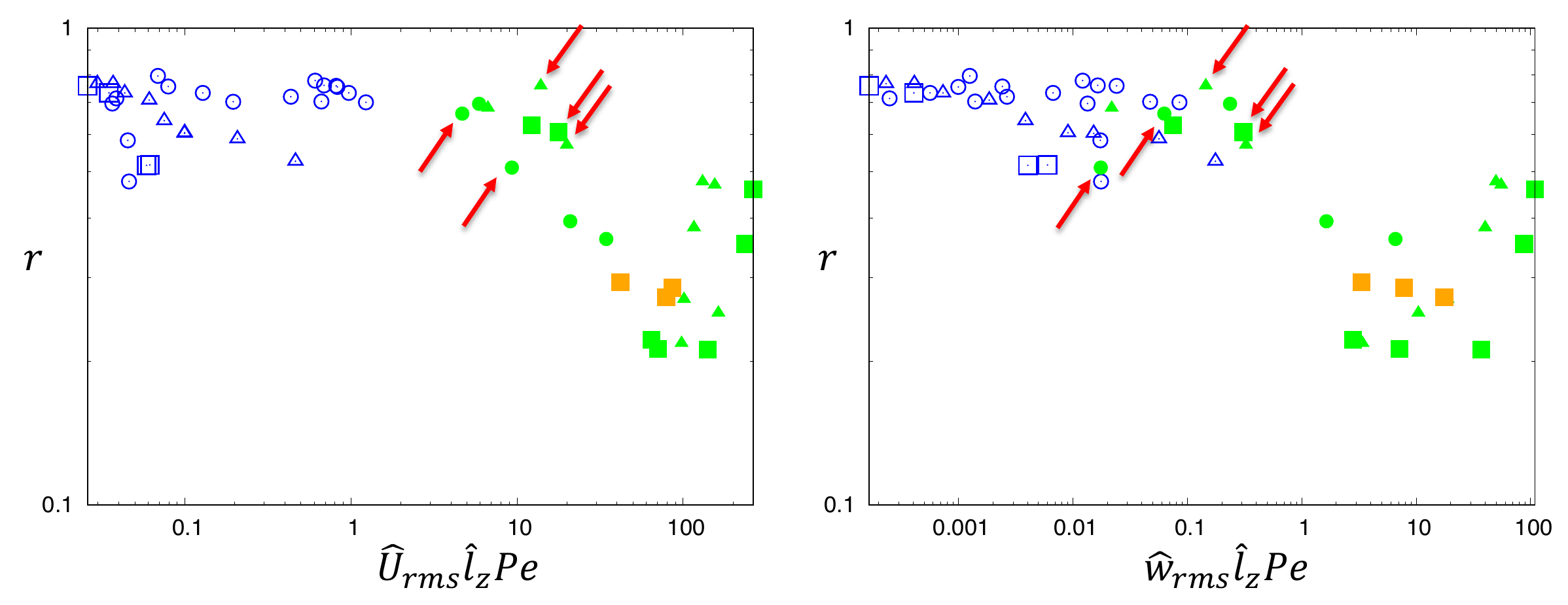}
\caption{The diagnostic quantity $r$ (see equation \ref{eq:r}) plotted against two different possible versions of the turbulent P\'eclet number: on the left, defined as $Pe_t =\hat U_{rms} \hat l_z Pe$, and on the right, defined as $Pe_t = \hat w_{rms} \hat l_z Pe$.  We see that the second option better distinguishes between simulations that satisfy the LPN approximation ($r \simeq 1$) and simulations that do not ($r \simeq 0.2-0.4$). The red arrows point to the simulations circled in Figure \ref{fig:HighPe_results}, whose properties are the same as low P\'eclet number simulations with the same value of $Re$ and $BPe$.}
\label{fig:rvsPerms}
\end{figure}

 Taking the analysis of \citet{Lignieres1999} at face value, one should therefore compare $r$ to a turbulent P\'eclet number based on the rms velocity of the fluid 
 \begin{equation}
 \hat U_{rms} = \sqrt{ \hat u_{rms}^2+ \hat v_{rms}^2 + \hat w_{rms}^2},
 \label{eq:Urms}
 \end{equation}
 and the vertical eddy scale, $\hat l_z$. The comparison is shown in Figure \ref{fig:rvsPerms}a, using the same symbol style as in Figure \ref{fig:HighPe_results}. We see that $r \simeq 1$ for all the low $Pe$ runs (blue symbols), which is expected since they also have $Pe_t = \hat U_{rms} \hat l_z Pe \ll 1$. At the other end of the scale, we see that for many of the high $Pe$ runs (green and orange symbols), for which $Pe_t \gg 1$, $r$ drops to values between 0.2 and 0.4, again as expected. However, we see a group of points for values of $Pe_t \simeq 10$ (which is greater than one) that nevertheless have $r \simeq 1$. The points marked with a red arrow are the same as those circled in Figure \ref{fig:HighPe_results}, whose properties are almost identical to those of low P\'eclet number simulations. This suggests that $Pe_t = \hat U_{rms} \hat l_z Pe$ is not the relevant bifurcation parameter for low P\'eclet number dynamics. 
 
 To correct this problem, we show in Figure \ref{fig:rvsPerms}b  the same data plotted this time against $Pe_t$ defined as 
 \begin{equation}
  Pe_t = \hat w_{rms} \hat l_z Pe.
  \label{eq:pet}
 \end{equation}
  We now see a much clearer partitioning between data with $Pe_t \ll 1$ that has $r \simeq 1$, and data with $Pe_t \gg 1$ which has $r \simeq 0.2-0.4$. All the simulations which had a red arrow have now moved to the low $Pe_t$ clump.
  
The fact that the definition of $Pe_t$ based on $\hat w_{rms}$ is a better choice than the one that uses $\hat U_{rms}$ is fairly surprising, since the derivation of \citet{Lignieres1999} of the LPN approximation clearly uses the latter rather than the former (and is indeed the correct formal way of deriving it). A possible way of understanding why this may be the case is to consider the horizontal average of the temperature equation, 
 \begin{equation}
 \frac{\partial \overline {\hat T}}{\partial t} + \frac{\partial}{\partial z} \overline{\hat w \hat T} = \frac{1}{Pe} \frac{\partial^2 \overline {\hat T}}{\partial z^2},
\end{equation}
where the overbar denotes a horizontal average. Following standard derivations, we have used incompressibility to write $\hat {\bf u} \cdot \nabla \hat T = \nabla \cdot (\hat {\bf u} \hat T)$ and the divergence theorem together with horizontal periodicity to reduce this term to the vertical derivative of the temperature flux. Also note that the horizontal average of $\hat w$ vanishes for mass conservation. For the convective flux to be much smaller than the diffusive flux, we therefore need 
\begin{equation}
\overline{\hat w \hat T}  \ll  \frac{1}{Pe} \frac{\partial \overline {\hat T}}{\partial z} , 
 \end{equation}
 which can be approximated as $\hat w_{rms} \hat T_{rms} \ll Pe^{-1} T_{rms} \hat l^{-1}_z$ to get 
 \begin{equation}
Pe_t =  \hat w_{rms} \hat l_z Pe \ll 1,
 \end{equation}
 as required.
 
 \subsubsection{High P\'eclet number dynamics} 

\begin{figure}[h]
\epsscale{0.9}
\plotone{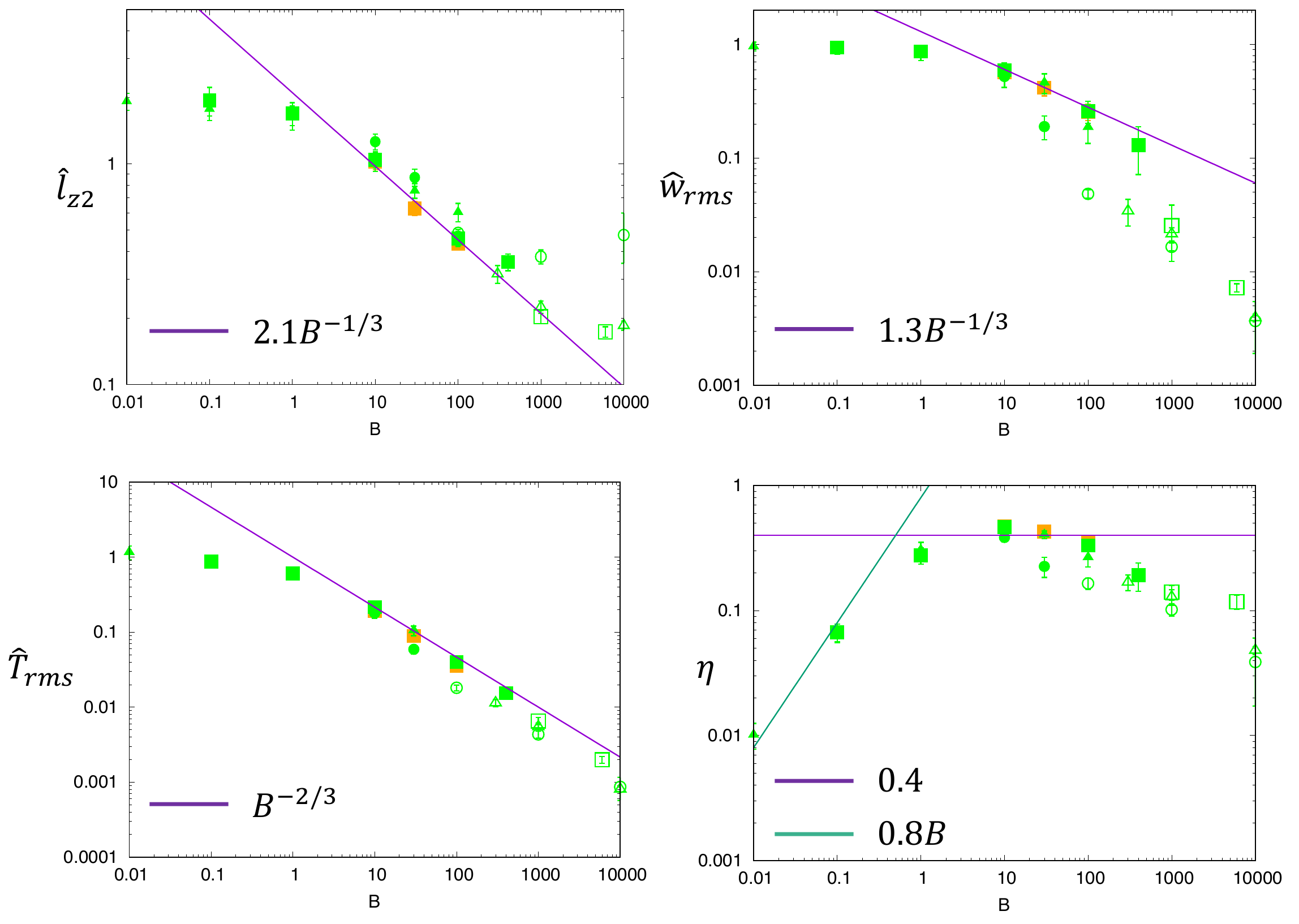}
\caption{From top left to bottom right: $\hat l_{z2}$, $\hat w_{rms}$, $\hat T_{rms}$ and $\eta$ as functions of $B$. In each quadrant, open symbols represent simulations which have $Pe_t \le 1$, while filled symbols have $Pe_t > 1$ (see equation \ref{eq:pet}); green symbols represent simulations at $Pr = 0.1$ and orange symbols have $Pr = 0.05$. The shape/size of the symbol represents the Reynolds number: small circles for $Re = 100$, small triangles for $Re = 300$ and large squares for $Re = 600$. The straight lines represent fits to the data in the stratified turbulent regime discussed in Section \ref{sec:scalings}.}
\label{fig:HighPe_vsB}
\end{figure}

So far, we have established that the $Pe \gg 1$ simulations presented in Table 2 can be partitioned into (1) simulations with $Pe_t \ll 1$ that have the characteristics of low P\'eclet number flows, which are now relatively well understood thanks to the work of \citet{Copeal20} and (2) simulations with $Pe_t \gg 1$ that do not have the characteristics of low P\'eclet number flows. We now focus on attempting to understand the latter. To do so, we present on Figure \ref{fig:HighPe_vsB} the same data as in Figure \ref{fig:HighPe_results}, but this time against $B$ instead of $BPe$. We have also removed the data for low $Pe$ (blue points and red points), and identify the high $Pe$ but $Pe_t < 1$ data with open symbols instead of filled symbols. Finally, for reasons explained in Appendix B, we have dropped the original definition of the vertical eddy scale $\hat l_z$ in favor of $\hat l_{z2}$, measured as 
\begin{equation}
\hat l_{z2} = \frac{\hat l'_z }{ 0.38}   \mbox{   where }  A_w(\hat l'_z,t) = 0.5 A_w(0,t) ,
\end{equation}
(i.e. where $\hat l_z'$ is the width of the autocorrelation function at half maximum). With this new definition, $\hat l_{z2}$ is close to the originally defined lengthscale for most simulations (see Appendix B), but is  more robust and less variable in time than $\hat l_z$. 
 
We see that presenting the data against $B$ causes it to collapse quite well into one universal curve for all the runs that are not at low $Pe_t$ (i.e. for all the filled symbols, except those that lie closest to the transition $Pe_t = 1$). This is expected: if indeed both $Pe_t \gg 1$ and $Re_t = Pe_t/Pr \gg 1$, one may anticipate all diagnostics of the flow to become independent of both parameters (though a weak dependence on their ratio $Pr$ remains possible). In the weakly stratified limit (which corresponds to $B \ll 1$), we see that, as in the low $Pe$ simulations of  \citet{Copeal20}, $\hat l_{z2}$ and $\hat w_{rms}$ tend to constants of order unity. By contrast, however, we now have $\hat T_{rms} = O(1)$ instead of $\hat T_{rms} = O(Pe)$, and $\eta \propto B$ instead of $\eta \propto BPe$.  Once $B$ exceeds unity, stratification becomes important and we enter the stratified turbulent regime. Empirically, we find that $\eta \simeq 0.4$, as in \citet{Copeal20}; this appears to be a general characteristic of the mixing efficiency in low Prandtl number flows. We also find that $\hat l_{z2} \sim \hat w_{rms} \sim B^{-1/3}$, and $\hat T_{rms} \sim B^{-2/3}$. The stratified turbulent regime appears to end as $Pe_t$ drops below unity, at which point the system satisfies the low P\'eclet number approximation and is well described by the theory of  \citet{Copeal20} (see more on this below). In the following section, we present a theory that explains the empirical scalings found in both the weakly stratified regime and in the stratified turbulent regime.

\subsubsection{Scaling laws}
\label{sec:scalings}

In what follows, we use $\hat l_z$ generically to denote a vertical lengthscale, and reserve $\hat l_{z2}$ for the lengthscale measured in the simulations (see Appendix B). 
In the weakly stratified regime, with the non-dimensionalization selected, we expect the eddies to be relatively isotropic with a dominant scale of order unity, and all three components of the velocity should also be of order unity \citep[see also][]{Copeal20}. Figure \ref{fig:HighPe_vsB} confirms that this is indeed the case for $\hat l_{z2}$ and $\hat w_{rms}$ when $B <1$. Furthermore, since the diffusion term in the temperature equation is negligible (this being a high P\'eclet number flow), we expect a balance between $\hat {\bf u} \cdot \nabla \hat T$ and $\hat w$, so that 
\begin{equation}
\hat w_{rms} \sim O(1) \sim \frac{ \hat w_{rms} \hat T_{rms} }{\hat l_z} \sim \hat T_{rms} ,
\end{equation}
implying that $\hat T_{rms}$ must also be of order one, as seen in Figure \ref{fig:HighPe_vsB}. Finally, noting that the denominator in $\eta(t)$ (see equation \ref{eq:etadef}) must always be $O(1)$ since $\hat u \sim O(1)$ then
\begin{equation}
\eta \simeq - B \langle \hat w \hat T \rangle \sim B \hat w_{rms}  \hat T_{rms} \sim B.
\end{equation}
as seen in the data.

In the regime of stratified turbulence, on the other hand, we expect the stratification term to become relevant. This does not directly affect the horizontal component of the momentum equation, so  we still expect to have $\hat p_{rms} \sim \hat u_{rms}^2 \sim O(1)$. In the vertical component of the momentum equation, on the other hand, the buoyancy term becomes important, and from hydrostatic balance (namely $\partial \hat p/\partial z \simeq B\hat T$) we obtain 
\begin{equation}
\frac{\hat p_{rms} }{\hat l_z} \sim \hat l_z^{-1} \sim B\hat T_{rms} .
\end{equation}
Meanwhile in the temperature equation we still expect the same balance as in the weakly stratified case (namely $\hat {\bf u} \cdot \nabla \hat T \sim \hat w$), but this time the eddy scale $\hat l_x \sim\hat  l_z$ is not necessarily $O(1)$, so 
\begin{equation} 
\frac{\hat u_{rms} \hat T_{rms} }{\hat l_z} \sim \frac{\hat T_{rms} }{\hat l_z}  \sim \hat w_{rms}  .
\end{equation}
Finally, as in the low $Pe$ analysis of \citet{Copeal20}, we assume that this regime is {\it defined} by a constant $\eta \sim O(1)$, which implies that 
\begin{equation}
B \hat w_{rms} \hat T_{rms} \sim O(1).
\end{equation}
Combining these three estimates we get 
\begin{equation}
\hat w_{rms} \sim \hat l_z \sim B^{-1/3}  \mbox{  and  } \hat T_{rms} \sim B^{-2/3} ,
\end{equation}
which is consistent with the observed scalings at intermediate values of $B$ (i.e. $B \gg 1$ but small enough for $Pe_t \gg 1$ to hold). A fit to the data can help constrain the prefactors and reveals that
\begin{equation}
\hat l_z \simeq 2.1 B^{-1/3}, \quad \hat T_{rms} \simeq B^{-2/3}, \mbox{  and  } \hat w_{rms} \simeq 1.3B^{-1/3}.
\end{equation}
These fits to the regime of stratified turbulence are shown as purple lines in Figure \ref{fig:HighPe_vsB}.

\subsubsection{Mixed layers and $U/N$ scaling}
\label{sec:mixed}

While the scalings derived above are quite plausible in the light of the supporting data, they 
are strikingly different from what is commonly discussed and observed in high Reynolds number / high P\'eclet number flows in geophysics, where $Pr > 1$. 
There, it is well known that the strongly stratified turbulence can intermittently drive the formation of localized mixed layers with reduced stratification 
separated by thinner interfaces with stronger stratification. The layers have a vertical scale of $l_z \sim (U/N)$, where $U$ here is more generally the rms velocity of horizontal flows, and $L$ is their horizontal scale  \citep[see, e.g.][]{Park1994,Holford1999,BillantChomaz2000,Brethouwer2007,Oglethorpe2013,Zhou2019}. When written in terms of the non-dimensionalization adopted in this work, the layer heights should therefore scale as $\sim B^{-1/2}$. We clearly do not see this scaling here. This is surprising since when $Pe \gg 1$ and $Re \gg 1$, the flow dynamics should be relatively independent of the microscopic parameters $\nu$ and $\kappa_T$ (and therefore of their ratio), so the theoretical arguments put forward to explain the formation of layers on a scale $B^{-1/2}$ in geophysical flows \citep{Brethouwer2007} should still apply here. This raises the question of whether thermally mixed layers on the scale $B^{-1/2}$ actually do exist in our simulations, but cannot be identified with the current method used to measure the vertical eddy scale. 
 
\begin{figure}[h]
\epsscale{0.9}
\plotone{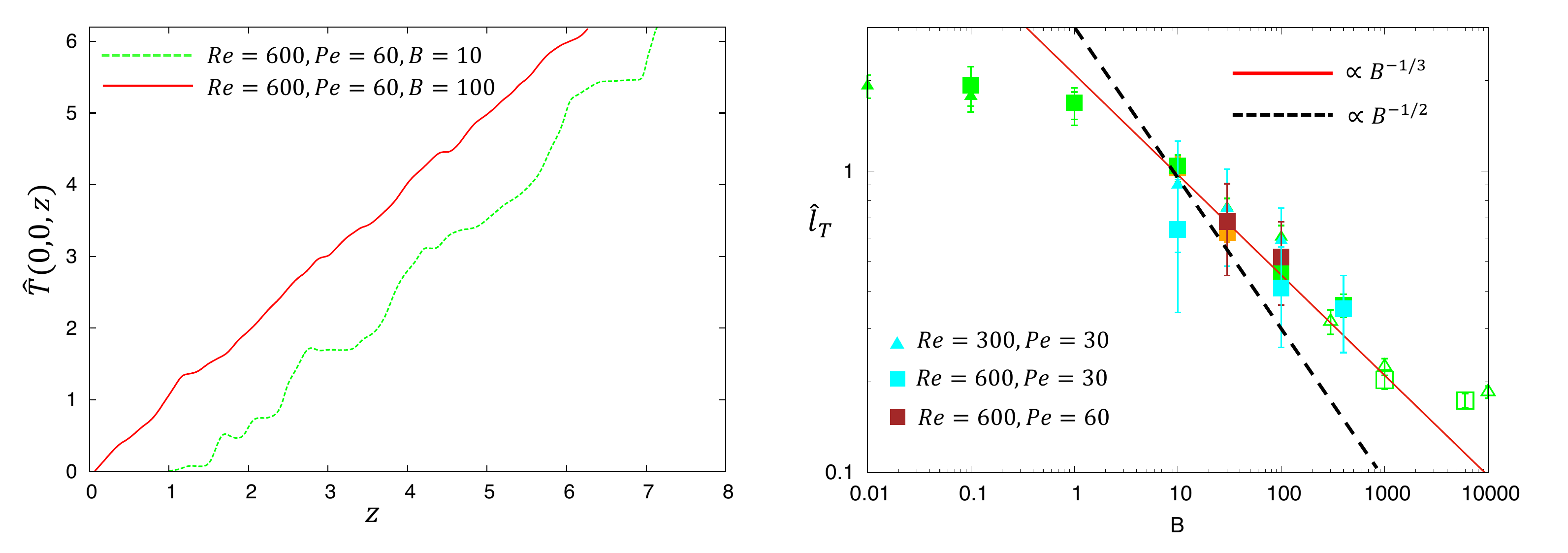}
\caption{Left: Selected profiles of the total temperature (e.g. $z+\hat T(x,y,z)$), for two simulations at $Re = 600$, $Pe = 60$, and $B = 10$ (green dashed line) and $B = 100$ (red solid line), respectively. The two profiles are offset horizontally for ease of visualization. Note the presence of steps, that each correspond to a mixed layer. The steps clearly become smaller as $B$ increases. Right: Height of the mixed layers $\hat l_T$ (cyan and brown symbols), measured using the method described in Appendix B, and compared with the vertical eddy scale (green and orange symbols), as a function of $B$. Symbols for the eddy scale are the same as in Figure \ref{fig:HighPe_vsB}. }
\label{fig:layers}
\end{figure}

Inspection of instantaneous temperature profiles (e.g. $\hat T(0,0,z)$ at different instants in time) in various simulations do reveal the presence of locally mixed layers, at least in the region of parameter space associated with stratified turbulence. This is shown in Figure \ref{fig:layers}a. These local inversions of the temperature gradient become smaller and rarer as $B$ increases, and for values of $B$ where $Pe_t \ll 1$, the temperature fluctuations are too small to cause any change in the background stratification. We have measured the scale $\hat l_T$ of these locally mixed regions, using the method described in Appendix B (note that for very small values of $B$ where temperature behaves more like a passive scalar, the temperature profiles are too variable to clearly identify layers, so we ignore them here). The results are presented in Figure \ref{fig:layers}b, and clearly show that these mixed layers have approximately the same width as the vertical eddy scale measured using the autocorrelation function -- in other words, each individual overturning event can be attributed to a single strong eddy, that locally mixes the background stratification. We find no evidence for a scaling law with $\hat l_T \sim B^{-1/2}$, as one might have expected. We are therefore forced to conclude that the behavior of low Prandtl number stratified turbulence is fundamentally different from that of high Prandtl number stratified turbulence, and that scalings typically associated with the latter do not apply here.  

\subsubsection{Transition to low P\'eclet number dynamics}
\label{sec:LPNtransition}

Using the new scaling laws derived in Section \ref{sec:scalings}, and the fact that the transition to low P\'eclet number dynamics occurs when $Pe_t$ drops below one, we predict that it should take place (roughly) when 
\begin{equation}
2.7  Pe B^{-2/3} < 1 \Leftrightarrow B > B_\kappa \simeq (2.7 Pe)^{3/2}.
\end{equation} 
For runs with $Pe = 10$, $30$ and $60$, respectively, the transition should take place around $B_\kappa \simeq 140$, $730$, and $2060$, respectively.  
This corresponds roughly to what we see in the data (within a factor of about 2). 

Note that this transition from stratified turbulence with high P\'eclet number dynamics to low P\'eclet number dynamics is unique to the low Prandtl number limit. Indeed, another way in which the stratified turbulence regime could break down is in the limit where viscosity becomes important. This happens when the viscous term in the horizontal component of the momentum equation grows to be of the same order as the other terms (which are all of order unity), namely when 
\begin{equation}
Re^{-1} \frac{\hat u_{rms}}{\hat l_z^2} \sim  O(1) \Leftrightarrow \hat l_z \sim Re^{-1/2}.
\end{equation}
With $\hat l_z \simeq 2.1 B^{-1/3}$ in the stratified turbulence regime, this transition would happen at the critical value 
\begin{equation}
B_\nu = 2.1^3 Re^{3/2}.
\end{equation}
However, since $Re \gg Pe$ when $Pr \ll 1$, we always have $B_\nu \gg B_\kappa$ so viscosity does not affect the transition from high P\'eclet number stratified turbulence to low P\'eclet number stratified turbulence. 

\begin{figure}[h!]
\epsscale{0.5}
\plotone{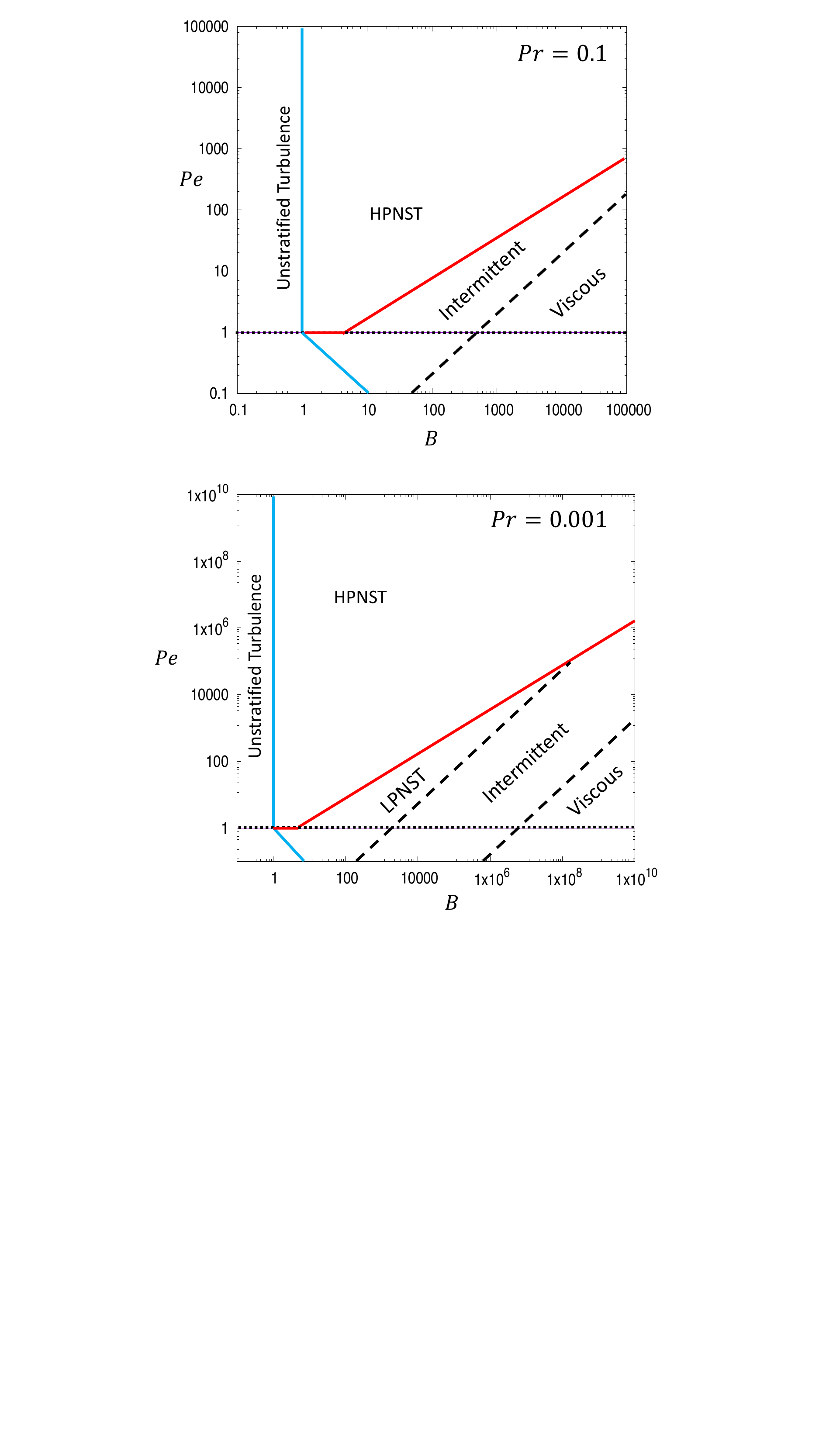}
\caption{\small Evolution of the regime diagram as the Prandtl number decreases from $Pr = 0.1$ (top, DNS value) to $Pr = 0.001$ (bottom). In both plots, the blue line marks the transition from unstratified (or weakly stratified) to strongly stratified turbulence, and the red line approximately marks the transition between high and low P\'eclet number dynamics; the inclined portion of the red line is the line $Pe = B^{2/3} / 2.7$, or equivalently, $B = B_\kappa$ or $Pe_t = 1$. Above the red line and to the right of the blue line the turbulence is in the high P\'eclet number stratified turbulence regime (HPNST) discussed in Section \ref{sec:scalings}. Below the red line, the dynamics satisfy the LPN approximation, and can fall into the three possible regimes identified by \citet{Copeal20}: low P\'eclet number stratified turbulence regime (LPNST), intermittent regime and viscous regime (see Section \ref{sec:Cope}).}
\label{fig:regimes}
\end{figure}

Once $Pe_t$ drops below one (or equivalently, when $B$ exceeds $B_\kappa$), then the flow is governed by the LPN approximation (see equation \ref{eq:LPN}). We know from the work of \citet{Copeal20} that the 
dominant dynamics can be classified into three possible regimes (ignoring the unstratified regime, which is not relevant for these strongly stratified shear flows): the low P\'eclet number stratified turbulence regime (LPNST), when $1 \ll BPe \ll 0.0016 Re^2$, the intermittent regime, for  $0.0016 Re^2 \ll BPe \ll 5 Re^2$, and the viscous regime, for $BPe \gg 5 Re^2$. Which of these three regimes the system transitions into as $B$ begins to exceed $B_\kappa$ therefore depends on $Pr$ and $Pe$, as illustrated in Figure \ref{fig:regimes}a. If $Pr$ is closer to one (e.g. $Pr = 0.1$, as in the DNSs presented here), then the flow transitions directly from high P\'eclet number stratified turbulence to the low P\'eclet number intermittent regime unless $Pe$ is very large. As $Pr$ decreases down toward stellar values, however, the flow can transition from high P\'eclet number stratified turbulence (HPNST) to low P\'eclet number stratified turbulence for intermediate values of $Pe$ (see Figures \ref{fig:regimes}b and \ref{fig:regimes2}). To see this numerically would require DNSs at the following parameters at least: $Pr = 0.001$, $Pe = 10$, and $Re = 10^4$, which is presently outside of the range achievable by the PADDI code. 

\section{Discussion}
\label{sec:disc}

\subsection{Summary, implications and discussion}

In this work, we have used DNSs to examine turbulent mixing in horizontal shear flows driven by a body force with amplitude $F_0$ and characteristic lengthscale $L$, in the distinguished stellar limit where the Prandtl number $Pr = \nu/\kappa_T$ is low, where both the {\it outer scale} Reynolds number $Re = UL / \nu$ and P\'eclet number $Pe = UL /\kappa_T$ are high, and where the stratification parameter $B = N^2 L^2 / U^2$ is high. Here, $N$ is the Brunt-V\"ais\"al\"a frequency, and in the model $U$ is a characteristic amplitude of the horizontal flow obtained by assuming a balance between the forcing and the Reynolds stress in the horizontal (see equation \ref{eq:Udef}). In a star, however, $U/L$ would simply be the observed mean horizontal shear. We have found that the resulting turbulent dynamics of these shear flows can be divided into two categories, depending on the turbulent P\'eclet number $Pe_t = w_{rms} l_z / \kappa_T = \hat w_{rms} \hat l_z Pe$, where $w_{rms}$ is the typical vertical velocity of turbulent eddies, and $l_z$ is their vertical scale. Note that hatted quantities are non-dimensional (see Section \ref{sec:model}), while non-hatted quantities are dimensional. 

In the more weakly stratified cases (but still with $B \gg 1$, as expected in stars), the dominant turbulent eddies and their vertical velocities are relatively large, so $Pe_t \gg 1$. In that limit, we found that $\hat l_z \simeq 2.1 B^{-1/3}$ which implies, dimensionally, that 
\begin{equation}
l_z \simeq 2.1 \left(\frac{N^2L^2}{U^2}\right)^{-1/3} L.
\label{eq:lzdimscal}
\end{equation}
Meanwhile, we found that  $\hat w_{rms} \simeq 1.3 B^{-1/3}$ and $\hat T_{rms} \simeq B^{-2/3}$, which imply dimensionally that 
\begin{eqnarray} 
w_{rms} && \simeq 1.3 \left(\frac{N^2L^2}{U^2}\right)^{-1/3} U, \mbox{   and  }   \\
T_{rms} && \simeq \left(\frac{N^2L^2}{U^2}\right)^{-2/3} L \left( T_{0z} - T_{ad,z} \right) \simeq \left(\frac{N^2L^2}{U^2}\right)^{-2/3} L \frac{N^2}{g} T_m, 
\label{eq:wrmsdimscal}
\end{eqnarray}
 where $T_{0z}$ is the background temperature gradient, $T_{ad,z}$ is the adiabatic temperature gradient, $g$ is gravity and $T_m$ is the mean temperature of the region considered. Taken together, these imply a vertical turbulent diffusivity (for compositional mixing or momentum transport for instance) 
 \begin{equation}
 D_{turb} \sim w_{rms} l_z \simeq 2.7 \left(\frac{N^2L^2}{U^2}\right)^{-2/3} UL, 
 \end{equation}
 and a vertical temperature flux (recalling that $|\hat F_T| \simeq 0.25 \hat w_{rms} \hat T_{rms}$ in this limit, see Section \ref{eq:LPNdynamics}), 
 \begin{equation}
 F_T \simeq - 0.3  \left(\frac{N^2L^2}{U^2}\right)^{-1} UL \left( T_{0z} - T_{ad,z} \right).
 \label{eq:FTdimscal}
 \end{equation}
 Note that all of these scaling laws were obtained by analyzing two sets of DNSs, one for $Pr = 0.1$, and one for $Pr = 0.05$, both of which are much larger than the expected values of $Pr$ in stars (which would be closer to $Pr \sim 10^{-6}$ or even smaller), and not particularly well separated in parameter space from one another. As such, there is a reasonable possibility that the prefactors in the estimates obtained have a weak dependence on $Pr$ (possibly logarithmic), leading to  uncertainties of order one in $l_z$, $w_{rms}$, $T_{rms}$, $D_{turb}$ and $F_T$. 
  
\begin{figure}[h]
\epsscale{0.5}
\plotone{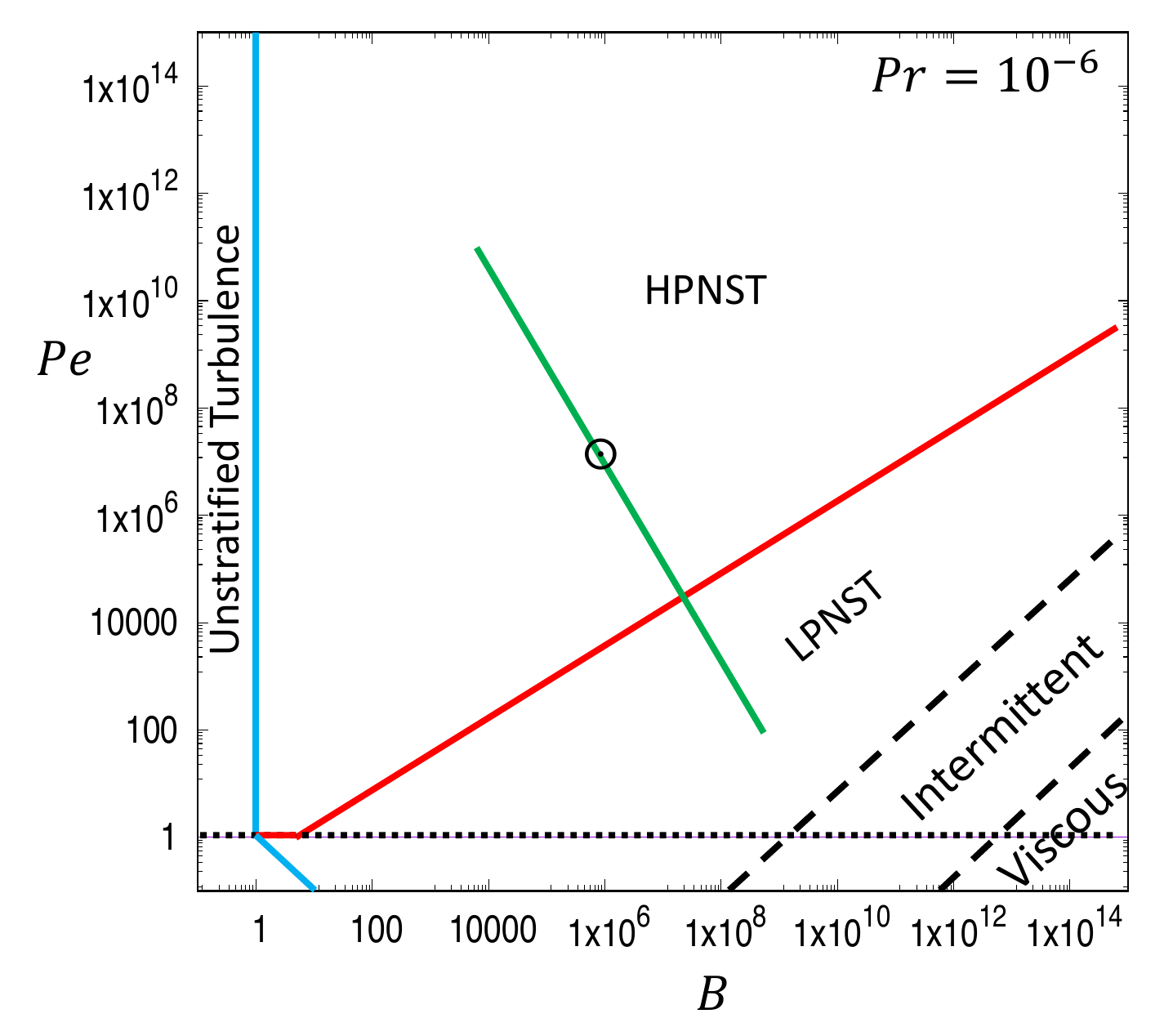}
\caption{\small As in Figure \ref{fig:regimes} but for a solar value of the Prandtl number. The $\odot$ symbol marks the approximate location of the bulk of the solar tachocline, and the green line shows how this position would change if $U$ varies (the point would go up along this line if $U$ increases, and down if $U$ decreases).}
\label{fig:regimes2}
\end{figure}

 As the stratification increases (i.e. $B$ increases holding everything else constant), both $l_z$ and $w_{rms}$ decrease, and so does the turbulent P\'eclet number $Pe_t$. When $Pe_t$ drops below one, the dynamics become thermally diffusive \citep[see][and Section \ref{sec:LPNtransition}]{Lignieres1999}. Assuming that the turbulence satisfies the scalings (\ref{eq:lzdimscal}) to (\ref{eq:FTdimscal}) prior to this diffusive transition, then the latter occurs when $B = B_{\kappa} = (2.7 Pe)^{3/2}$, independently of $Pr$ (see Section \ref{sec:LPNtransition}). For $B \gg B_\kappa$, the temperature equation satisfies the LPN approximation (\ref{eq:LPN}). As discovered by \citet{Copeal20} and summarized in Section \ref{sec:Cope}, there are various possible regimes the system could achieve in that case, depending on the respective values of the product $BPe$ and of $Re$ (low P\'eclet number stratified turbulence regime, intermittent regime, and viscous regime). A possible regime diagram for stellar values of the Prandtl number ($Pr \sim 10^{-6}$) is presented in Figure \ref{fig:regimes2}, showing both the diffusive transition, and the possible regimes achievable beyond the transition. Generally speaking, we see that for reasonable stellar values of $B$ and $Pe$ (see, e.g. equation \ref{eq:tachopars}), we can expect a simple transition from high P\'eclet to low P\'eclet stratified turbulence as $B$ increases.
 
 \subsection{Implications for the solar tachocline}

 As discussed in Section \ref{sec:intro}, the solar tachocline is the best-known example of a stellar shear layer that is located in a radiative zone, and has substantial horizontal shear. 
 Using values of $L \simeq r_{cz} = 5 \times 10^{10}$cm (where $r_{cz}$ is the radius of the base of the convection zone), $N \simeq 10^{-3}$s$^{-1}$, $U \sim r_{cz} \Delta \Omega \simeq 10^4$cm/s (where $\Delta \Omega \simeq 2 \times 10^{-7}$s$^{-1}$), $\nu \simeq 10$cm$^2$/s, and $\kappa_T \simeq 10^7$cm$^2$/s appropriate for the bulk of the solar tachocline, we 
get
\begin{equation}
Pr \simeq 10^{-6}, Re \simeq 10^{13}, Pe \simeq  10^{7} \mbox{  and  } B \simeq 10^6.
\label{eq:tachopars2}
\end{equation}
This point is shown in Figure \ref{fig:regimes2} and lies well within the high P\'eclet number stratified turbulence regime discussed above. 
 Combining (\ref{eq:lzdimscal}) to (\ref{eq:FTdimscal}) with (\ref{eq:tachopars2}) would imply that 
 \begin{equation}
l_z \simeq 2.1 \times 10^{-2} L,  w_{rms} \sim 1.3 \times 10^{-2} U, \mbox{   and   } T_{rms} \sim 10^{-4}  L \left( T_{0z} - T_{ad,z} \right). \label{eq:tachopred}
\end{equation} 
Since the width of the tachocline itself is {\it at most} of the order of a few percent of $r_{cz}$ this would appear to imply that it is only a few eddies wide (or less). 
We can also compute an estimate for the vertical turbulent diffusivity (of chemical species, or momentum) as
\begin{equation}
D_{turb,v} \sim 2.7 \times 10^{-4} UL  \sim O(10^{11}) \mbox{cm}^2\mbox{/s}.
\label{eq:Dturbv} 
\end{equation}
Estimating the horizontal turbulent diffusivity $D_{turb,h}$ from the model on the other hand is much more difficult, because the horizontal flow contains a vast range of energy-containing scales (from the scale of the jet and its large-scale meanders, to the scale $l_x \simeq l_y \simeq l_z$ of the turbulent eddies). As such, it is not clear whether one should estimate $D_{turb,h} \sim UL$, or $D_{turb,h} \sim U l_z$, or something else altogether. 

Nevertheless, we are now in a position to determine whether our turbulence model is consistent with the \citet{SpiegelZahn92} model of the tachocline, or not. 
\citet{SpiegelZahn92} demonstrated that, provided the tachocline is turbulent, and provided the turbulence is sufficiently anisotropic such that  
\begin{equation} 
\frac{D_{turb,h}}{D_{turb,v}} \gg \left( \frac{r_{cz}}{h} \right)^2, 
\end{equation}
then the tachocline thickness is related to $D_{turb,h}$ via
\begin{equation} 
h \simeq \left(\frac{\Omega_\odot}{N}\right)^{ 1/2} \left( \frac{\kappa_T}{D_{turb,h}}\right)^{1/4} r_{cz} 
\end{equation} 
(see their equation 5.19). For Spiegel \& Zahn's model to be self-consistent, we therefore need
\begin{equation}
\frac{h}{r_{cz}} \ll \left(\frac{\Omega_\odot}{N}\right)  \left( \frac{\kappa_T}{D_{turb,v}}\right)^{1/2}.
\label{eq:consistency}
\end{equation}
Assuming that our new turbulence model is indeed applicable to the solar tachocline, then (\ref{eq:Dturbv}) should hold. Substituting the observed values of all known quantities (see above, and also $\Omega_\odot \simeq 3 \times 10^{-6}$s$^{-1}$), (\ref{eq:consistency}) becomes 
\begin{equation}
\frac{h}{r_{cz}} \ll \left(\frac{\Omega_\odot}{N}\right)  \left( 2.7B^{-2/3}Pe\right)^{-1/2} \sim O(10^{-4}) .
\label{eq:consistency2}
\end{equation}
This is {\it not} consistent with our turbulence model\footnote{Whether this is consistent with observations or not remains to be determined -- observations can still only provide an upper limit on the tachocline thickness.}, where the height of the turbulent eddies is $O(10^{-2}) r_{cz} \gg O(10^{-4}) r_{cz}$ (see equation \ref{eq:tachopred}). In other words, it appears that the Spiegel \& Zahn model of the tachocline cannot be reconciled with our new model for stratified turbulence driven by horizontal shear flows. Possible resolutions of this inconsistency are discussed below.

Finally, from (\ref{eq:FTdimscal}) we find that the turbulent temperature flux is  
\begin{equation}
|F_T |  \sim O(10^4) \mbox{ K cm/s } , 
\end{equation}
using a value of $T_{0z} - T_{ad,z} = N^2 T_m / g \simeq  10^{-4}$K/cm with $g \simeq 5 \times 10^4$cm/s$^2$ and $T_m \simeq 2 \times 10^6$K. 
This is to be compared with the background diffusive temperature flux, which is equal to $-\kappa_T T_{0z} \sim 10^3$Kcm/s using $|T_{0z}| \simeq 10^{-4}$K/cm.  The ratio of the two is therefore of the order of 
\begin{equation}
\frac{|F_T|}{\kappa_T |T_{0z}|} \sim 10 , 
\end{equation} 
which would imply that the shear-induced turbulence could have a substantial effect on the heat transport in this region. Note that being located in a stably stratified region, $F_T < 0$, which would imply an inward turbulent heat flux. 
Again, this finding is not consistent with Spiegel \& Zahn's model of the tachocline, which assumes that the shear-induced turbulence does not affect the local stratification.

\subsection{Discussion}

The apparent contradictions between our numerical findings on stratified turbulence generated by horizontal shear flows and Spiegel \& Zahn's model of the tachocline \citep{SpiegelZahn92} strongly suggests that one or the other (or both) may not appropriately model the tachocline dynamics. If our new turbulence model is correct, then this calls for a completely new model of the solar tachocline, in which the turbulence is quite strong and able to modify the stratification below the convection zone. This would likely be observable using helioseismology.  If on the other hand Spiegel \& Zahn's model applies, then this would imply that our turbulence model is missing crucial elements that need to be accounted for to correctly capture the tachocline dynamics. There are several possibilities in which this could be the case.

For instance, it is important to bear in mind that our model predictions for the solar tachocline are predicated on the assumption that there is no other possible turbulent regime. However, this is just an assumption, and it is not impossible that a new regime could appear at very low Prandtl number, with its own set of scaling laws. If that were to be the case, then the functional dependence of $Pe_t$ on $Pe$ and $B$ could change, in which case the $Pe_t  =1$ line would move in parameter space from its present position. In other words, future work will be needed to confirm (or invalidate) the predictions made in this work when $Pr$ is in the stellar range.

More importantly, however, is the fact that our study currently neglects several important physical processes that are known to be present in stars and will likely impact the results, such as rotation, magnetic fields, vertical shear and the possibility of additional sources of stratification such as a gradient in chemical composition.


The most likely culprit is rotation. As discussed by \citet{Watson81} \citep[see also][]{Garaud01,Parkal2020}, rotation can stabilize a global latitudinal differential rotation pattern  (at least from the perspective of linear theory), so the 2D mode of horizontal shear instability that is always present in our simulations (and is crucial to the excitation of the turbulence) may disappear in rapidly rotating stars, or in stars that are weakly differentially rotating. In the solar tachocline, the shear appears to be marginally stable to horizontal shear instabilities, which could be interpreted as evidence that the tachocline is actually turbulent, and that the turbulence is transporting  potential vorticity to drive the system toward (but never quite reaching) marginal stability \citep[see][]{Garaud01}. Even if the flow is shear-unstable, however, rotation is likely to influence both 2D and 3D modes of instability \citep{Parkal2020}, therefore affecting the large-scale horizontal eddies and their horizontal transport properties. We can quantify this by estimating the Rossby number associated with the vertical component of the momentum equation (i.e. the ratio of the nonlinear terms to the Coriolis term). We find that it is the same as that associated with horizontal flows, and equal to 
\begin{equation}
Ro_v \sim \frac{|{\bf u} \cdot \nabla w|}{|{\bf \Omega}_\odot \times {\bf u}|} \sim \frac{w_{rms}}{\Omega_\odot l_z} \sim \frac{B^{-1/3}U}{\Omega_\odot B^{-1/3} L} \sim Ro_h \sim  10^{-1} . 
\end{equation}
This implies that rotation will be important even on the smaller vertical scales associated with the eddies, and will likely modify the vertical momentum balance crucial to the turbulence scalings derived in Section \ref{sec:scalings}. Further study of the effect of rotation on the results presented in this paper is therefore crucial to a better understanding of the solar tachocline in particular, and other stars in general.

Coherent horizontal magnetic fields (such as a large scale toroidal field that is likely present in the tachocline) could also stabilize the standard 2D hydrodynamic mode of instability, but would in turn drive alternative types of magnetohydrodynamic modes \citep[e.g.][]{GilmanFox1997,Cally2001}, that would behave quite differently from the large-scale meanders that arise in our model. Furthermore, since magnetic fields are generated on all scales by the turbulence, they will likely modify the vertical momentum balance, with similarly crucial consequences on the scalings derived. Again, a further study of the effect of magnetic fields will be required before the model can be reliably applied to the Sun and other stars. 

Beyond the addition of rotation and magnetic fields, the model will also need to account for compositional stratification and vertical shear. Indeed, the solar tachocline is a region that is subject to both horizontal and vertical shear (rather than horizontal shear alone, as studied here), and it is not clear whether the added vertical shear would affect our results or not. Finally, 
a compositional (rather than thermal) stratification would significantly change the results discussed here as well. This is because the compositional diffusivity $\kappa_C$ is typically smaller than the kinematic diffusivity by a factor of 10 or so in stars, so the equivalent Prandtl number $\nu/\kappa_C$ would be larger than one instead of being small. In that case, results from the geophysical literature are more likely to apply (in particular those obtained for thermally stratified water, where $Pr \simeq 10$). 

Despite the enormous task still lying ahead, however, the present study provides a first numerical look at the possible nature of turbulence in stably stratified regions of stars undergoing horizontal shear, even if it might not necessarily 
apply to the solar tachocline. It is quite clear that horizontal shear flows have the potential to cause substantial vertical mixing in stars, which ought to be taken into account in stellar evolution models from here on.

\acknowledgments

P. G. acknowledges the support of NSF AST 1814327. The simulations were performed using the PADDI code kindly provided by S. Stellmach, on the Comet supercomputer of the NSF XSEDE infrastructure, and on the LUX supercomputer at UCSC, funded by NSF MRI AST 1828315. P.G. thanks Brant Robertson (UCSC), Joshua Sonstroem (UCSC), and the XSEDE support staff for their technical help. Figure 1 was created using VisIt. VisIt is supported by the Department of Energy with funding from the Advanced Simulation and Computing Program and the Scientific Discovery through Advanced Computing Program. P.G. also thanks L. Cope and C. P. Caulfield for their insight into stratified turbulence that helped interpret some of the model results. This work is dedicated to Edward Spiegel and Jean-Paul Zahn, whose legacy continues to inspire us.

\appendix 
\section*{Appendix A: Stationary vs. non-stationary runs}
This Appendix briefly presents some of the raw data obtained from the simulations, and illustrates the extraction procedure. It also discusses the issue encountered for the few runs at high values of $B$, in which a statistically 
stationary state has not yet been reached. 

\begin{figure}[h]
\epsscale{0.9}
\plotone{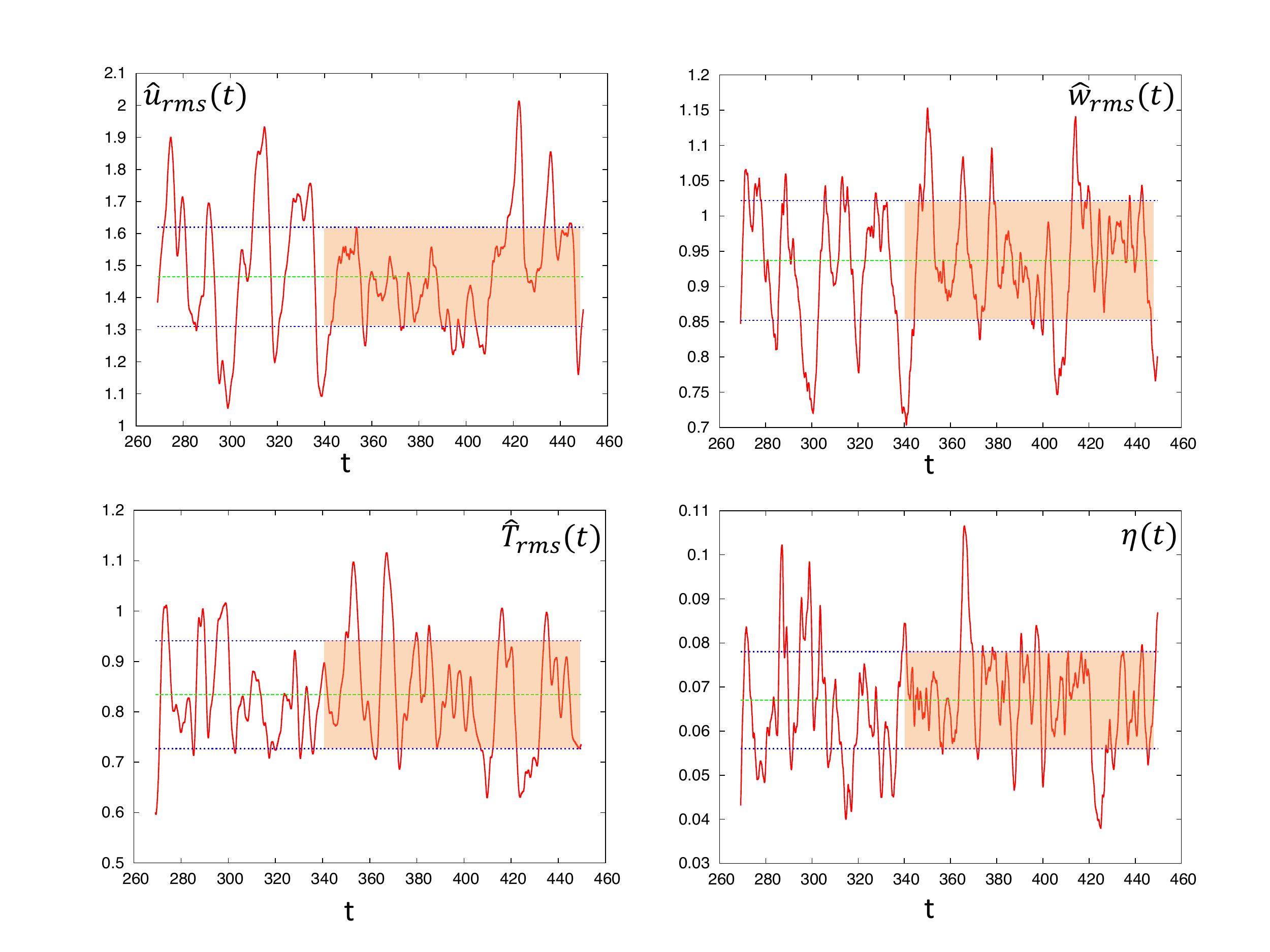}
\caption{Time evolution of $\hat u_{rms}(t)$, $\hat w_{rms}(t)$, $\hat T_{rms}(t)$ and $\eta(t)$ for a simulation with $Re = 300$, $Pe = 30$, $B = 0.1$, restarted from a run at $Re = 300$, $Pe = 30$, $B = 1$. The orange shaded box marks the time during which the system is deemed to be statistically stationary. The green line is the measured average, and the two blue lines show the average plus and minus one standard deviation.}
\label{fig:AppendixFig1}
\end{figure}

For all simulations, we measure volume averaged quantities such as $\hat u_{rms}(t)$, $\hat w_{rms}(t)$, $\hat T_{rms}(t)$ and $\eta(t)$ (see Section \ref{sec:dataextract}). In all cases, quantities associated with vertical motions settle into a statistically stationary state very rapidly, while $\hat u_{rms}(t)$ (and $\hat v_{rms}(t)$, to some extent) often take longer to reach this state. A simulation is therefore deemed to have achieved such a state if $\hat u_{rms}(t)$ appears to be statistically stationary for a interval of duration $\Delta t = 100$ or more (which corresponds roughly to 100 turnover times of the horizontal eddies, since the latter have both size and velocity $\sim O(1)$ in the non-dimensionalization selected). In most cases presented in Table \ref{tab:runs} such a state has been achieved, and the time averages of, e.g. $\hat u_{rms}(t)$, $\hat w_{rms}(t)$, $\hat T_{rms}(t)$ and $\eta(t)$ are then measured and reported, together with their rms variability around the average. Figure \ref{fig:AppendixFig1} shows an example of a simulation at $Re = 300$, $Pe = 30$, $B = 0,1$ that appears to have reached a statistically stationary state. For each of the quantities plotted, the green line is the mean measured between $t = 340$ and $t = 450$, while the blue lines are one rms above and one rms below that average. 

\begin{figure}[h]
\epsscale{0.9}
\plotone{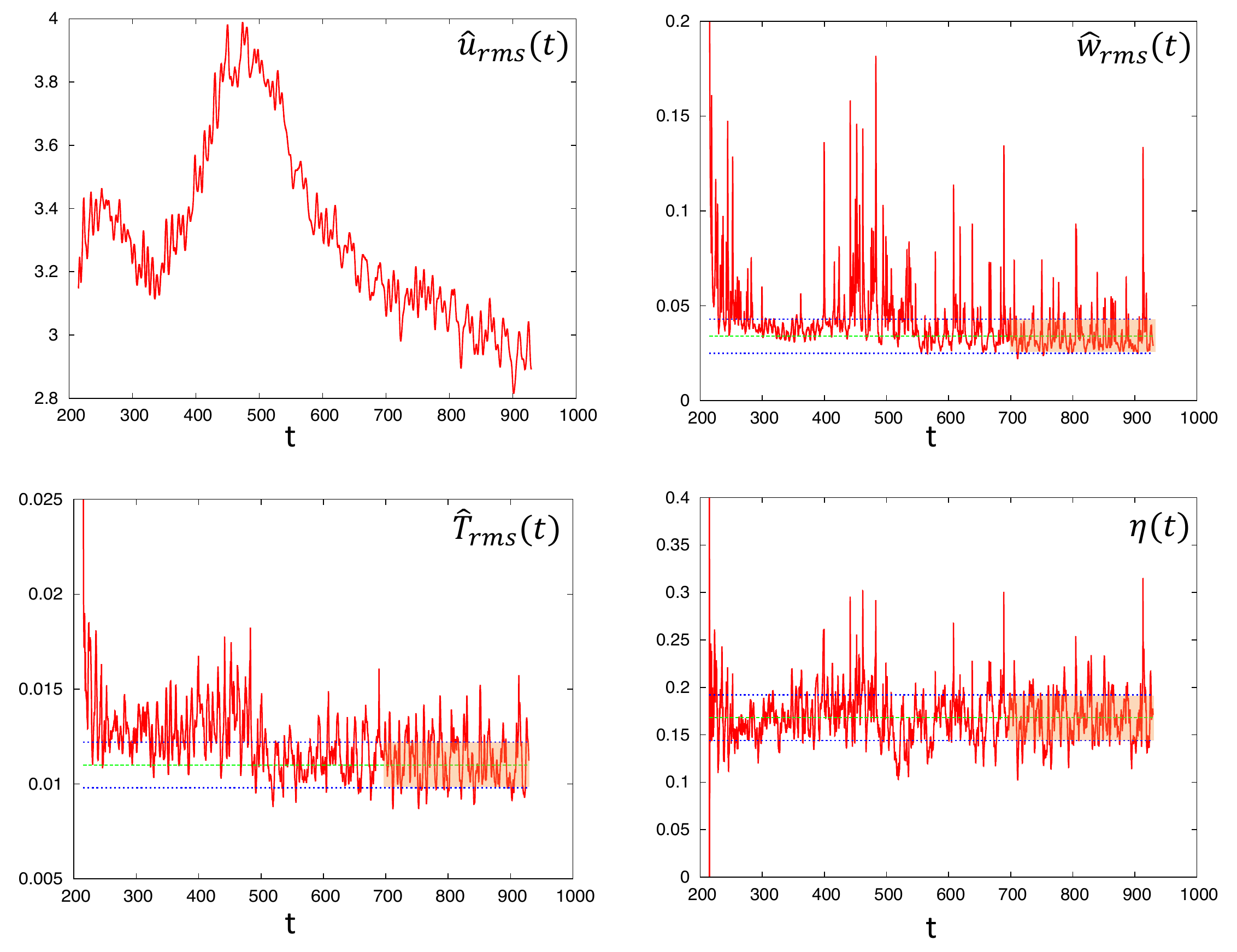}
\caption{Time evolution of $\hat u_{rms}(t)$, $\hat w_{rms}(t)$, $\hat T_{rms}(t)$ and $\eta(t)$ for a simulation with $Re = 300$, $Pe = 30$, $B = 300$, restarted from a run at $Re = 300$, $Pe = 30$, $B = 100$. The orange shaded box marks the time during which the system is deemed to be statistically stationary in terms of $\hat w_{rms}(t)$, $\hat T_{rms}(t)$ and $\eta(t)$, even though  $\hat u_{rms}(t)$ is still evolving. The green line is the measured average, and the two blue lines show the average plus and minus one standard deviation.}
\label{fig:AppendixFig2}
\end{figure}

By contrast, a few simulations at high values of $B$ do not appear to have reached such a state yet, despite considerable integration times. This is the case for example of the $Re = 300$, $Pe = 30$, $B = 300$ run, shown in Figure \ref{fig:AppendixFig2}. We see, however, that quantities associated with vertical motions have settled into a statistically stationary state, which appears to start roughly around $t = 700$. The averages of  $\hat w_{rms}(t)$, $\hat T_{rms}(t)$ and $\eta(t)$ were therefore measured in the time interval between $t = 700$ and $t = 900$. The average of $\hat u_{rms}(t)$ has also been measured, but should not be viewed as statistically stationary.

\section*{Appendix B: Lengthscale measurements}

\begin{figure}[h]
\epsscale{0.9}
\plotone{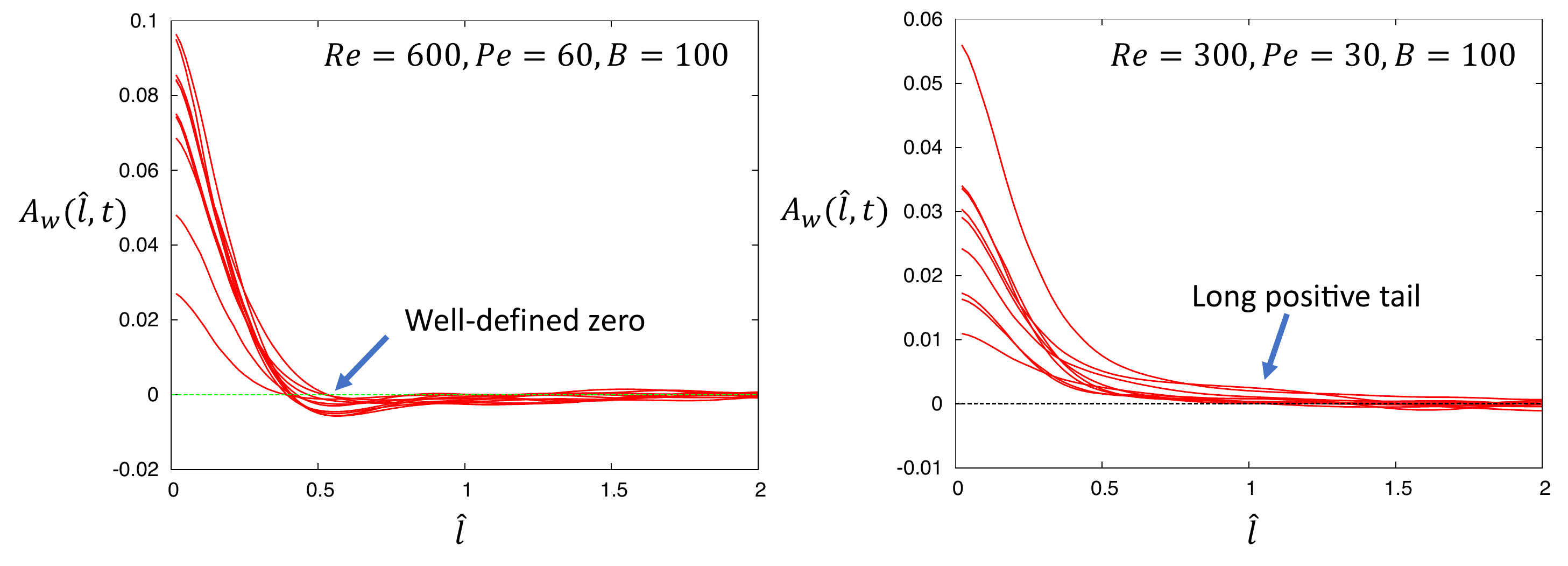}
\caption{Comparison between a normal case, where the first zero of the autocorrelation function $A_w(\hat l,t)$ is well defined (left), to an abnormal case, where there is instead a long positive tail (right). In both figures there are nine lines, corresponding each to a graph of $A_w(\hat l,t)$ as a function of $\hat l$ at a specific point in time taken during the statistically stationary phase.}
\label{fig:Aw}
\end{figure}

As discussed in the main text, \citet{Copeal20} defined the vertical lengthscale $\hat l_z(t)$ of turbulent eddies at a given instant in time as the first zero of the autocorrelation function $A_w(\hat l,t)$  (see equation \ref{eq:Aw}). In most cases, this definition works very well as the zero is well defined and fairy stationary in time. However, in a few of the high P\'eclet number runs presented in Section \ref{sec:num}, we have found that $\hat l_z(t)$ varies widely with time, because $A_w(\hat l,t)$ has a long positive but weak amplitude tail whose first zero exhibits wide excursions. The difference between the normal and abnormal behavior of $A_w(\hat l,t)$ is illustrated in Figure \ref{fig:Aw}. 

\begin{figure}[h]
\epsscale{1}
\plotone{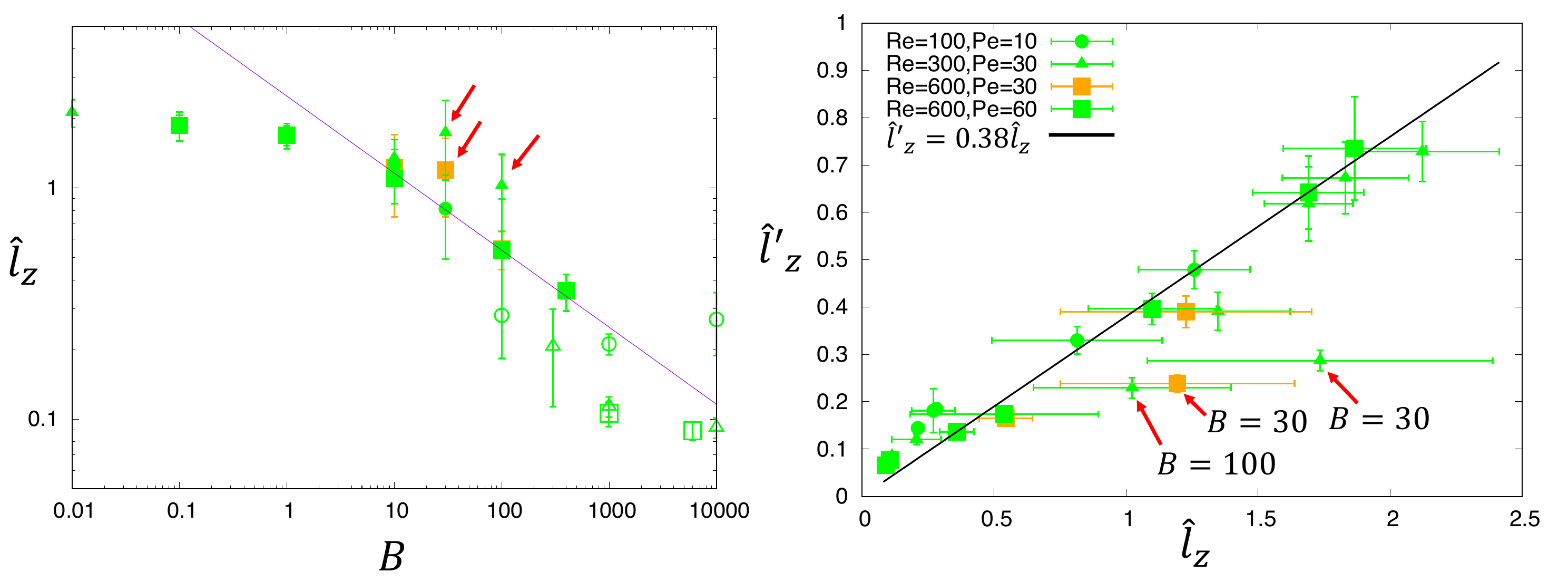}
\caption{Left: As in Figure \ref{fig:HighPe_vsB} in the main text, but showing $\hat l_z$ vs. $B$ instead of $\hat l_{z2}$ vs. $B$. The legend is on the right-side plot, and the purple line is the line $2.5 B^{-1/3}$. Note the three outliers, marked by the red arrows. Right: Comparison between $\hat l'_z $  and $\hat l_z$. Aside from three points marked by the red arrows (which correspond to the same simulations highlighted on the left), the data shows that $\hat l'_z$ is indeed roughly proportional to $\hat l_z$, with $\hat l'_z \simeq 0.38\hat l_z$ (black line).}
\label{fig:AppendixFig3}
\end{figure}

Inspection of the data revealed that the simulations for which the abnormal behavior is most pronounced are for $Re = 300$, $Pe = 30$, $B = 100$ (shown in Figure \ref{fig:Aw}), $Re = 300$, $Pe = 30$, $B = 30$, and $Re = 600$, $Pe = 30$ and $B = 30$. Interestingly, these are precisely the simulations which appear to be outliers when plotting $\hat l_z$ vs $B$ (see red arrows on Figure \ref{fig:AppendixFig3}a, which is the equivalent of Figure \ref{fig:HighPe_vsB} in the main text but with $\hat l_z$ instead of $\hat l_{z2}$). This strongly suggests that using the first zero of the autocorrelation function may not be a universally good estimate of $\hat l_z$ for the high P\'eclet number simulations. 

Figure \ref{fig:Aw} shows that in both normal and abnormal cases the function $A_w(\hat l,t)$ has a well defined core structure that can be used to create an alternative definition for the eddy size $\hat l_z$. We therefore define a new lengthscale $\hat l'_z(t)$ such that 
\begin{equation}
A_w(\hat l'_z,t) = 0.5A_w(0,t)
\end{equation}
(i.e. the width at half maximum), and as usual, take the time average of $\hat l'_z(t)$ during the statistically stationary state. Naturally, we expect $\hat l'_z < \hat l_z$ by continuity of $A_w(\hat l,t)$. We also expect that in most normal cases, $\hat l_z$ and $\hat l'_z$ should follow something close to a simple linear relationship, with $\hat l'_z$ proportional to $\hat l_z$ (that relationship would be exact if $A_w$ were a linear function of $\hat l$). We have measured $\hat l'_z $ for all simulations presented in Table \ref{tab:runs}, and plot the two lengthscales against one another in Figure \ref{fig:AppendixFig3}b. We see that overall, $\hat l'_z \simeq 0.38 \hat l_z$, except for the same three abnormal simulations that appear as outliers in the plot (marked as red arrows). We therefore adopt a new definition of $\hat l_{z2} = \hat l'_z / 0.38$ in the rest of the paper, to ensure that (other than the abnormal cases), $\hat l_{z2}$ is as close as possible to the originally-defined $\hat l_z$. 

\begin{figure}[h]
\epsscale{0.5}
\plotone{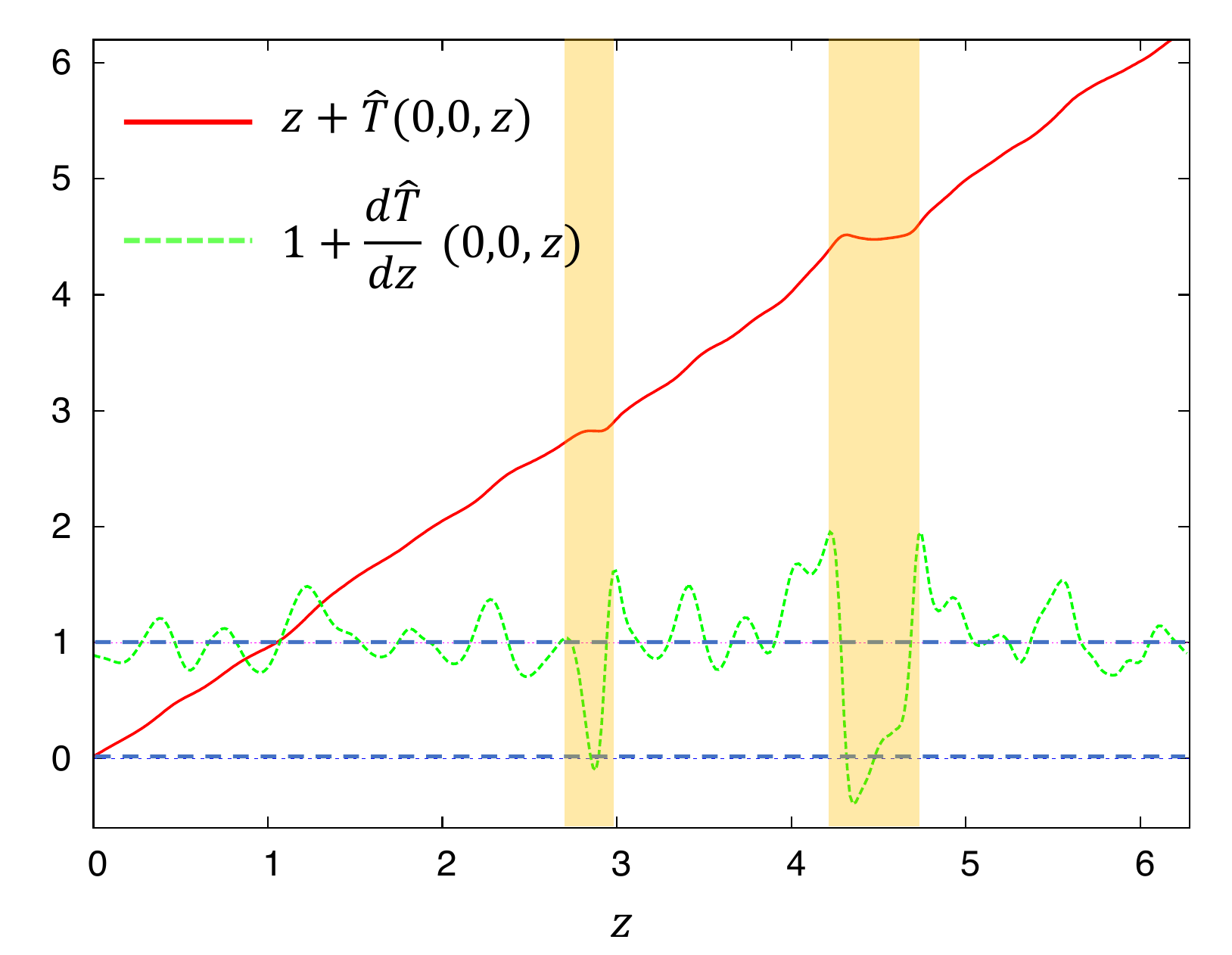}
\caption{Illustration of the method used to identify a thermally mixed region in the flow, from the simulation at $Re = 600$, $Pe=60$ and $B = 100$. Using the total temperature profile at a given position $(x,y)$ in the domain (red solid line), we compute the gradient (green dashed line). Minima below zero represent a mixed region, whose width is computed as the distance between two local maxima on either side whose value lies above one. The two mixed regions for this particular profile are shown as the shaded orange regions.}
\label{fig:MeasurelT}
\end{figure}

Finally, as discussed in the main text (see Section \ref{sec:mixed}), we also measured the vertical scale of thermally mixed layers $\hat l_T$ for simulations in the stratified turbulent regime (for $Pe_t \gg 1$). To do so, we looked at individual profiles $\hat T(x,y,z,t)$ for all $(x,y)$ points at selected instants in time where the full fields are available. We then constructed the total temperature $z + \hat T(x,y,z,t)$, and its gradient, $1+ d \hat T(x,y,z,t) / dz$. We identified all local minima and maxima of this gradient. A region is deemed to be thermally mixed if the minimum of $1+ d \hat T(x,y,z,t) / dz$ lies below zero; the corresponding width this region is then computed as the distance between the two nearest local maxima whose value is greater than 1 bracketing this minimum. An example of the procedure applied to a profile from the simulation at $Re = 600$, $Pe=60$ and $B = 100$ is shown in Figure \ref{fig:MeasurelT}.
 The procedure is repeated for all available profiles and the lengthscale $\hat l_T$ is then computed as the average width of all mixed layers identified. 
 
\bibliographystyle{aasjournal}

\providecommand{\noopsort}[1]{}\providecommand{\singleletter}[1]{#1}%


\end{document}